\documentclass[a4paper,11pt]{article}
\usepackage{jinstpub} 
\usepackage{orcidlink}

\title{\boldmath Particle detection performance and Geant4 simulation with low-cost CMOS technology}

\author[a]{M. ~Bonnett Del Alamo\,\orcidlink{0000-0002-1011-8411}}
\author[a]{R. ~Helaconde\,\orcidlink{0000-0001-7076-1575}}
\author[a]{C. ~Soncco\,\orcidlink{0000-0002-7330-0405}}
\author[a]{J. ~Bazo\,\orcidlink{0000-0001-9148-9101}}
\author[a]{A.M. ~Gago\,\orcidlink{0000-0002-0019-9692}}

\affiliation[a]{Secci\'on F\'isica, Departamento de Ciencias, Pontificia Universidad Cat\'olica del Per\'u \\
            {Av. Universitaria 1801}, 
            {Lima},
            {15088},
            {Per\'u}
            }

\emailAdd{mbonnett@pucp.edu.pe}

\abstract{
We evaluate the performance of an Omnivision OV5647 CMOS image sensor (5 Mp) for detecting radiation from Sr90 and Cs137 sources. Our experimental setup uses a Raspberry Pi 3 mini-computer for data acquisition, with image processing using Python and OpenCV libraries. We specify the necessary settings to convert a standard camera into a particle detector sensitive to electrons and photons, including a two-step background filtering procedure. In addition, we implement the first detailed Geant4 simulation that describes the layered geometry and material composition of a commercial CMOS sensor along with the radioactive sources. To enhance the simulation, we include an algorithm for charge diffusion and conversion of the energy deposited by electrons and photons into ADC counts. Our measurements are presented in terms of cluster size, the maximum ADC signal per cluster, and the number of clusters as a function of distance. We find a good agreement between the experimental data and simulation for all these observables, and we can reproduce the correlation between cluster size and maximum ADC signal per cluster. Thus, this simulation, cross-checked with data, can be used to test the feasibility of further particle detection ideas without the need to implement an experimental setup. However, the sensor has limited primary energy resolution and is thus unable to distinguish between different radioactive sources. Nevertheless, given the accurate measurement of energy deposition, the sensor, once calibrated, is suitable for dosimetric measurements of source activities.
}


\arxivnumber{} 

\begin{document}
\maketitle
\flushbottom

\section{Introduction}
\label{sec:intro}
Imaging sensors based on metal oxide semiconductors (CMOS), designed to take photographs, have been widely used to detect particles such as gamma rays, electrons, alphas, etc \cite{PEREZ2016171, 7237371, 8645967, Lipovetzky2020}.
One low-cost commercial CMOS sensor that has been used for this purpose is the 5 Mp OmniVision OV5647 camera, which has shown good results in particle detection, such as discriminating between alpha and non-alpha particles by identifying and extracting ionization events \cite{8645967} and photon imaging using fluorescence X-rays and gamma rays \cite{Lipovetzky2020}.

In this work, we define a straightforward experimental setup to detect electrons and photons from radioactive sources (Sr-90 and Cs-137) using the OV5647 sensor enclosed within a dark box. The setup includes a Raspberry Pi 3b for data acquisition. We provide a detailed description of the camera settings, configured using the Picamera libraries, to ensure stable and sensitive measurements. In addition, we outline a procedure to effectively eliminate background noise, ensuring the detection of pure signals from radioactive sources, identifying clusters of pixels corresponding to particle events.

To reproduce the experimental data, we implement, for the first time, a detailed simulation of a commercial sensor. Our simulation is divided into two main steps. First, we use Geant4 to model the geometry and materials of the sensor and radioactive sources, yielding the energy deposited in the pixel matrix. In the second step, we employ a custom algorithm to simulate charge diffusion, and to convert energy deposition into ADC counts.

This paper is divided as follows: in Section \ref{exp_setup} we describe the experimental setup, the characteristics of the OV5647 CMOS sensor, and the radioactive sources. Section \ref{dat_acqui} describes data collection and processing with the help of the Picamera libraries. Section \ref{geant4_sim} explains how we perform the simulation of the sensor and radioactive sources in Geant4. Section \ref{dat_analysis} describes data processing with the help of OpenCV libraries, the methods used to reduce background noise, and the search for clusters representing particle tracks in the images. In Section \ref{resul} we show the results of comparing the simulation with the experimental data for different parameters related to the number of ADC counts and clusters. We also study the behavior of the clusters with distance and the possibility of differentiating between radioactive sources. Finally, in Section \ref{conclu} we give our conclusions. 

\section{Experimental setup} \label{exp_setup}

\begin{figure}[htbp]
\centering
\includegraphics[width=0.8\textwidth]{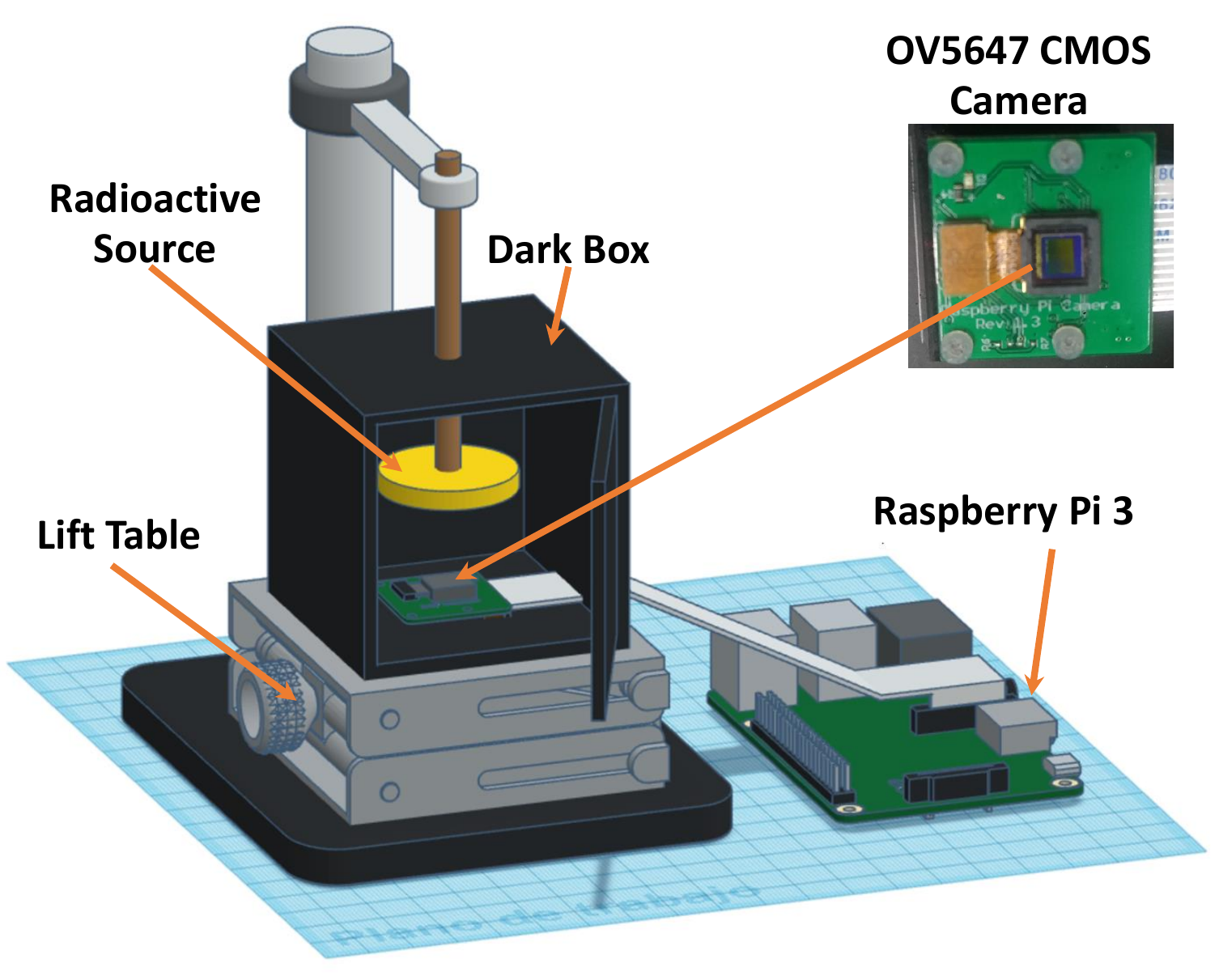}
\caption{Experimental setup to capture 10-bit images (the CMOS sensor, the radioactive source, and the Raspberry Pi 3 are to scale, while the other devices are not).} 
\label{fig:assembly}
\end{figure}

To test the CMOS sensor for particle detection, we designed the experimental configuration shown in figure\ref{fig:assembly} (devices not to scale). In the lower part, we placed a lift table, which moves the sensor along the $z$ axis and has a resolution of $100 \pm 50$ $\mu$m, allowing us to control the distance between it and the radioactive source. 

On the lift table, we put an aluminum dark box (68.87$\pm0.01$ mm high with a squared base of side 88.06$\pm0.01$ mm), black painted and sealed with black PVC insulation tape to ensure it is light-tight. Inside the dark box, we placed a radioactive source fixed to a support inside the box. The detector (OmniVision OV5647 CMOS sensor \cite{datasheets_rasp_camera_WC}) is placed at the base of the dark box.

To capture the frames with the OV5647 camera, we used a Raspberry Pi 3b (i.e. a small, low-cost single-board computer \cite{datasheets_raspberry3B, raspberry} ). To avoid light leakage to a minimum, we connect the OV5647 camera via a flexible cable to the Raspberry Pi 3b.

\subsection{OmniVision OV5647 Sensor} \label{detect_cmos}

The OmniVision OV5647 sensor (5 Mp) is a low-cost camera module that can be used along with a Raspberry Pi. It has a resolution of 2592 $\times$ 1944 pixels, with a pixel pitch of 1.4 $\times$ 1.4 $\mu$m$^2$ and a 3.67 $\times$ 2.73 mm$^2$ active area \cite{datasheets_rasp_camera_WC}.
The cross-section of the OV5647 CMOS sensor, as measured in \cite {10.1002_cta.2502, Lipovetzky2020}, shows the following layers and their thickness from top to bottom: microlenses for each pixel made from PDMS (SiOC$_2$H$_6$) (0.735 $\mu$m), the photoresist (C$_{10}$H$_6$N$_2$O) color filter to enable RGB imaging (0.9 $\mu$m), a thin insulator layer of SiO$_2$ (0.225 $\mu$m) and the sensitive Si detection volume (2 $\mu$m).

CMOS sensor pixels can detect photons and electrons in different ways. The detection of visible light photons is based on the photoelectric effect, where photons create electron-hole pairs directly in silicon.  The detection of electrons and high-energy photons (X-rays or gamma rays) is based on material ionization. These high-energy photons interact with the material through other processes such as photoelectric absorption or Compton scattering, releasing electrons that lose their energy as they cross the material and are able to create electron-hole pairs. In both cases, electron-hole pairs are generated, which are then picked up by the sensor's electronics. The energy used to generate an electron-hole pair in Si is on average 3.6 eV for both photons and electrons \cite{8522044, BRIGIDA2004322, 8645967, SCHOLZE2000208, LECHNER1996206}.

The maximum theoretical number of electrons a CMOS sensor can store before becoming saturated is called the "Full Well Capacity" (FWC).  In contrast, during normal operation, the sensor's photodiode typically holds a lower charge to ensure linearity and reliable performance. This is referred to as the "Well Capacity" (WC), which is the number of electrons a sensor can store at a specific gain level \cite{8645967, LANE201229, Bronnimann2014, El_Gamal2005_WC}. The OV5647 sensor has an FWC of approximately 4500 electrons per pixel. However, the WC is slightly lower, around 4300 electrons, when using an analog gain of 8 or higher, as indicated in the datasheet \cite{datasheets_rasp_camera_WC, FWC_OV5647}.

To get the source as close to the sensor as possible, the optical lens of the OmniVision OV5647 camera was removed, exposing the pixel sensor as shown in figure \ref{fig:assembly}.

\subsection{Radioactive sources} \label{radio_source}

\begin{figure}[htbp]
\centering
\includegraphics[width=0.8\textwidth]{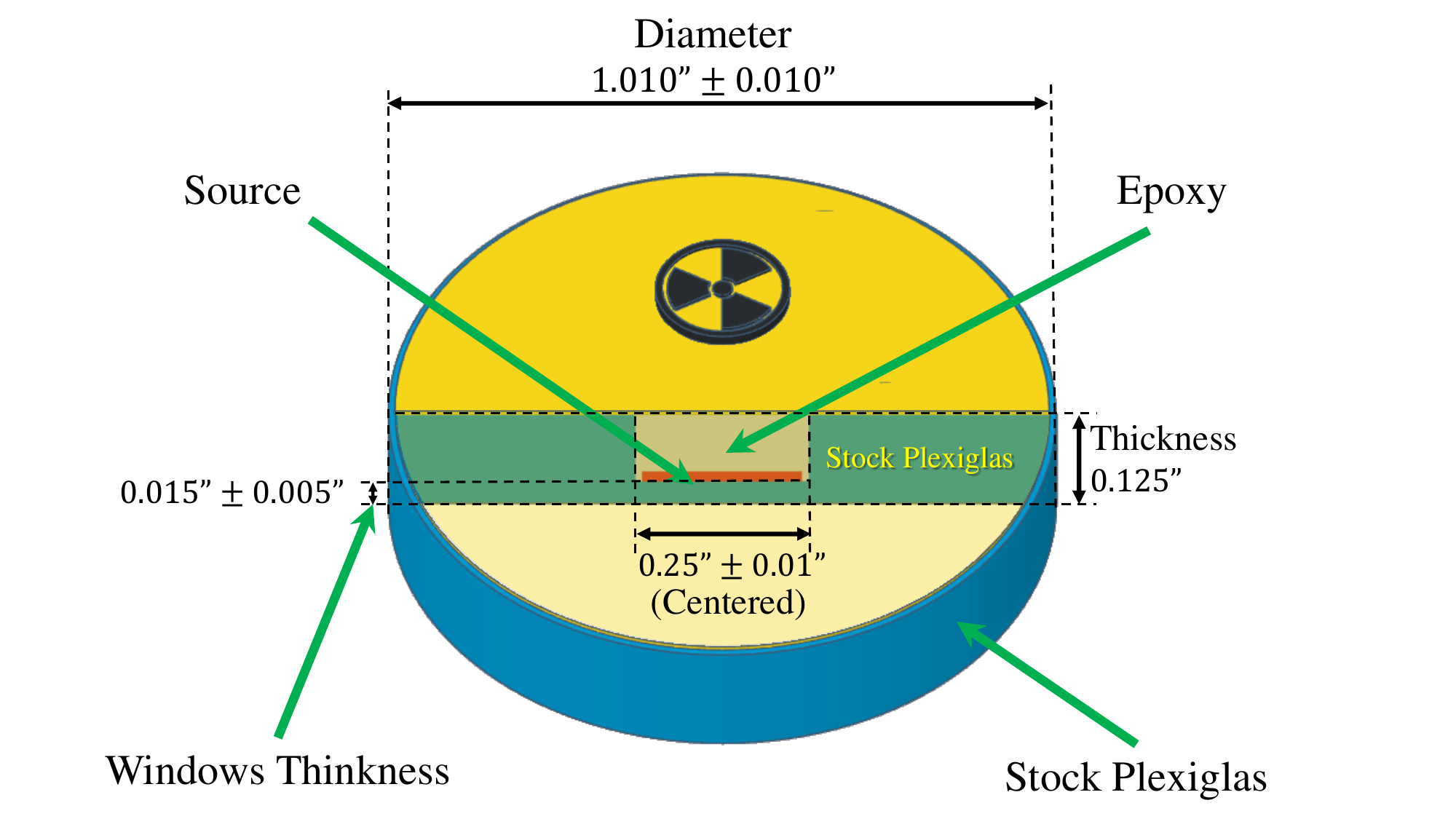}
\caption{Radioactive source geometry according to the information  provided by the manufacturer}
\label{fig:source_geometry}
\end{figure}

The radioactive sources that were used are Sr90 and Cs137 from \textit{Spectrum Techniques} in the form of cylindrical tablets. Inside an epoxy cylinder of 0.25" in diameter with a height of 0.11" is the radioactive source. This disk, as shown in figure \ref{fig:source_geometry}, is encased in a plexiglass cylinder of 1.010" diameter with a height of 0.125" \cite{spectrumtechniques}.

The Sr90 and Cs137 sources decay almost entirely by the emission of $\beta^-$ particles (electrons). The Sr90 source emits electrons with a maximum energy of 0.546 MeV, and the resulting Y90 isotopes also decay into electrons with a maximum energy of 2.28 MeV. Thus, Sr90 is technically a pure electron source, since gamma photon emission is negligible (\cite{nds_iaea_nuclides}, \cite{10.1093/oxfordjournals.rpd.a006705}). The characteristics provided by the manufacturer show that the Sr90 source was produced in July 2015 with an activity of 0.1 $\mu$Ci and a half-life of 28.8 years. With these data, we calculate the activity during data taking (i.e. October 2023) to be 3028 Bq.

The electron emission in the Cs137 decay is approximately 94.6\% with an energy of 0.514 MeV, producing Ba-137m, which emits a 0.662 MeV photon. Thus, Cs137 is a mixed source of electrons and photons \cite{ nds_iaea_nuclides}, \cite{10.1093/oxfordjournals.rpd.a006705}. The characteristics provided by the manufacturer show that the Cs137 source was produced in June 2015 with an activity of 0.25 $\mu$Ci and a half-life of 30.1 years. With these data, we calculate the activity during data taking (i.e. October 2023) to be 7622 Bq. 

\section{Data acquisition} \label{dat_acqui}

To record the data with our experimental setup, we use a Raspberry Pi3 and Picamera libraries \cite{picamera}. The latter allows us to configure the parameters of the OV5647 sensor. The regular use of the OV5647 is to take photographs, with that purpose various camera settings can change according to light conditions. Therefore, to ensure stability in the data acquisition we require to fix the following settings: 
\begin{itemize}
    \item Shutter speed = 0.5 seconds (i.e. the exposure time with which each frame will be taken). This time is enough to capture $\mathcal{O}(10^3)$ radioactive decays from the selected sources. 
    \item Image resolution = $2592\times1944$ pixels (5 Mp), which is the maximum resolution. 
    \item Analog gain = 8. This gives the maximum stable response of the camera without image distortion. 
    \item Digital gain = 1. This is an artificial gain performed by software set to its minimum value.
    \item White balance = 1. This value is equivalent to no color correction since we are not looking at visible light. 
\end{itemize}

To capture unprocessed images that use a color filter pattern, we used the 10-bit Bayer format of the Picamera library for the ADC counts, which does not compress the data. This format is defined by three color matrices (RGB) distributed in the CMOS sensor, so that 25\% of the pixels are red, 50\% green and 25\% blue \cite{picamera}. Since we are interested only in the intensity of each pixel and not its color, we add these three partial matrices to obtain a full matrix with the intensities in ADC values and save them in Python 'npz' format.

To obtain the background, we captured 1000 frames, equivalent to 500 seconds of actual exposure time, with the OV5647 inside the dark box without a radioactive source. The processing of each frame with the Raspberry Pi, including the matrix conversion and the data saving is approximately 11.5 seconds per frame, adding to a total of about 3 hours and 11 minutes of data taking. 

Then a radioactive source (Sr90 and Cs137, respectively) is placed on top of the OV5647 sensor at a distance of 0 mm between the source and the detector. In this configuration, the radioactive source with a circular area of 7.92 mm$^2$ illuminates approximately 74\% of the rectangular surface of the 10.02 mm$^2$ pixel sensor. The same amount of frames and duration was taken as in the background measurement. 
  
Additionally, with the help of the lift table (Figure\ref{fig:assembly}), the detector is moved away from the radioactive source, which is fixed by the support until there is a separation of 2 mm between them, and we take 100 frames each 0.5 seconds of data every 2 mm. This procedure is repeated every 2 mm until an 18 mm separation between the source and the detector is reached. The latter has been implemented to test the inverse square distance law based only on the Sr90 source measurements. 

Finally, we performed background measurements for three days at different times, showing a variation of 0.04\% to 0.08\% of active pixels, the average ADC value was between 11.5 and 11.7 with an error of 1.8, and their standard deviation was 4.2 $\pm$ 0.8. This variation is minimal, so we can consider the background to be stable over time.

\begin{figure}[htbp]
\centering
{\includegraphics[width=0.47\textwidth]{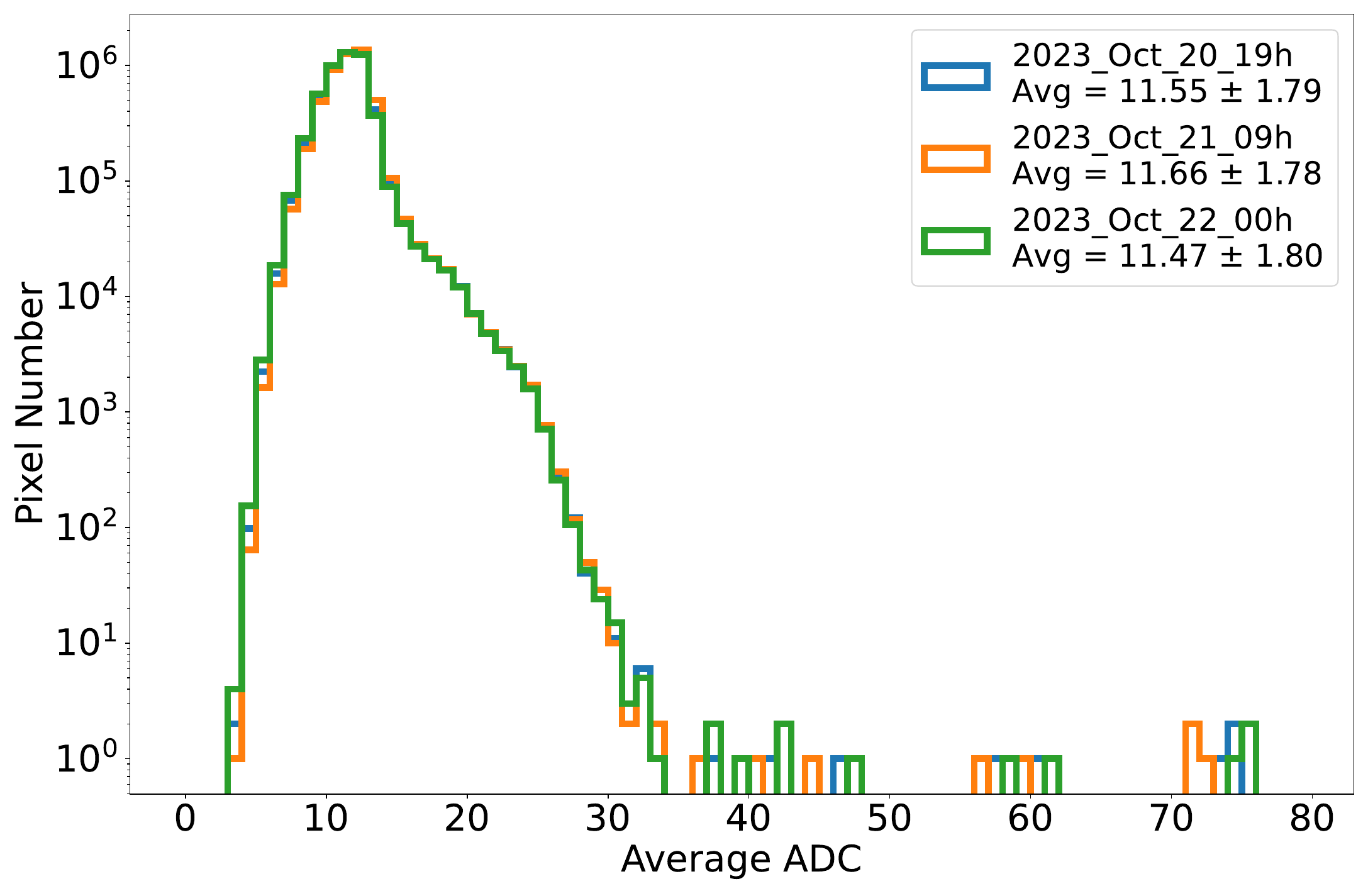}}
{\includegraphics[width=0.47\textwidth]{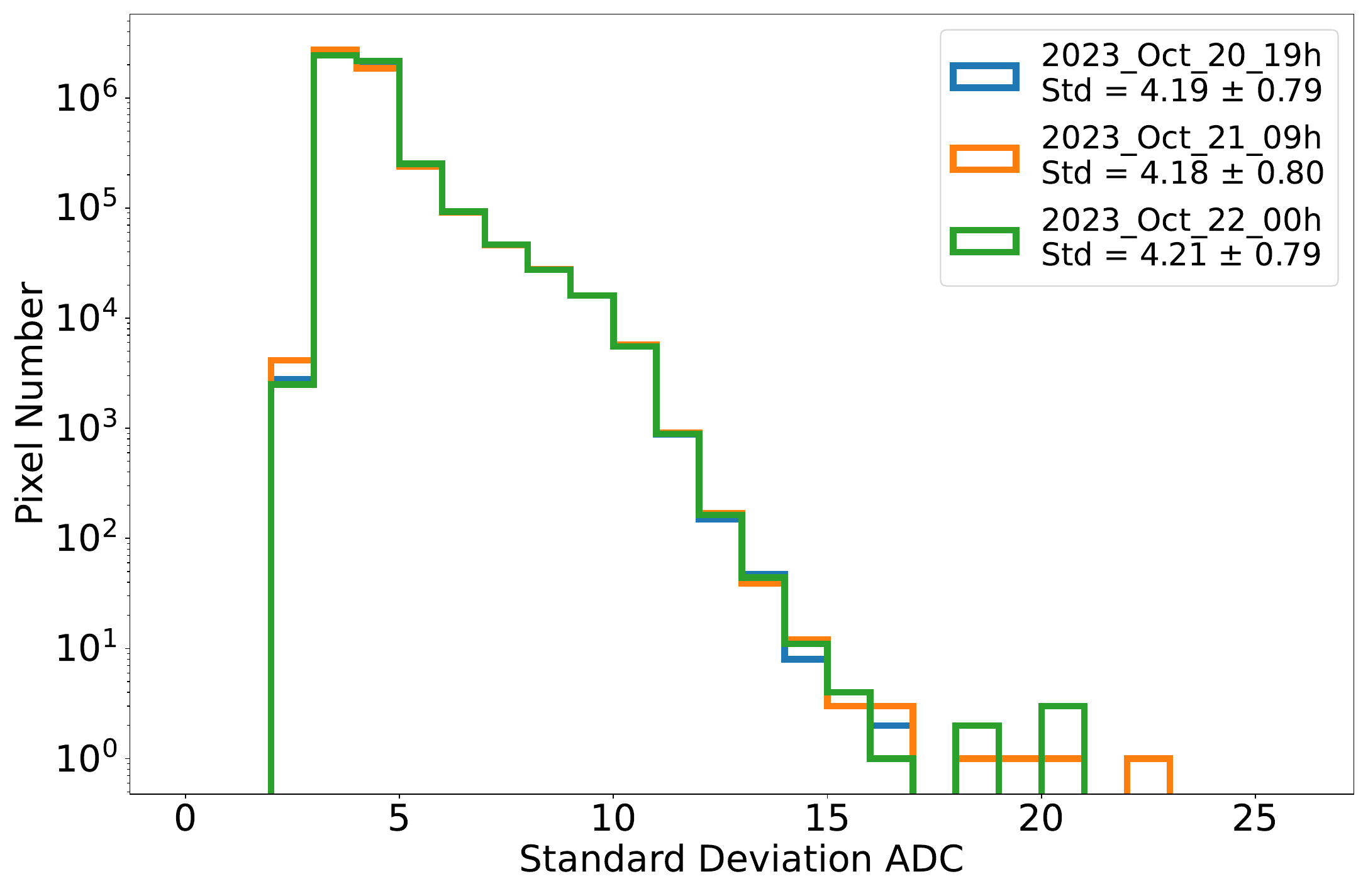}}
\caption{Distribution of the average (left) and standard deviation (right) for 1000 frames of the background noise of 3 datasets on different days.}
\label{fig:plot_hist_bg_avg_std_mean}
\end{figure}

Figure \ref{fig:plot_hist_bg_avg_std_mean}, shows the distribution of the average and standard deviation of the data taken for the background noise.

\section{Geant4 Simulation} \label{geant4_sim}

Our simulation is divided into two parts. First, we use Geant4 \cite{AGOSTINELLI2003250_geant4} to model the geometry and materials of the OmniVision OV5647 CMOS sensor and radioactive sources to reproduce the experimental data. In the second part, we employ a custom algorithm to simulate charge diffusion and convert energy deposition into ADC counts.

For the first part, the dimensions and materials of the CMOS sensor mentioned in Section \ref{detect_cmos} were used to define its geometry and composition. The geometry of the radioactive sources (i.e. Sr90 and Cs137) was simulated according to the specifications provided by the manufacturer, mentioned in Section \ref{radio_source}. This paper's simulation code and scripts can be found in the supplementary material \cite{scripts_data}.

To simulate in Geant4, we first describe the geometry using the “DetectorConstruction” class. We define the radioactive emission source through the “PrimaryGeneratorAction” class using the “G4GeneralParticleSource” class that allows the generation of particles from the whole volume of the source. We implement the “PhysicsList” class including radioactive decays, decays, and electromagnetic standard physics. 
To collect the information on the simulated events, such as deposit energy, position in the particle type sensor, etc.,  the “RunAction” and “EventAction” classes are used, as well as the “StackingAction” and “SteppingAction” classes. 

Finally, in order to change the distance between the source and the detector and other geometrical characteristics, as well as the type of radioactive source, the "DetectorMessenger" and "RunActionMessenger" classes were implemented, which allow us to create a macro to change these values.

Additionally, we made a Python script that creates the Geant4 macro for different distances and sources as well as converting the Geant4 data to '.npz' format to use our Python programs.

\begin{figure}[htbp]
\centering
{\includegraphics[width=0.65\textwidth]{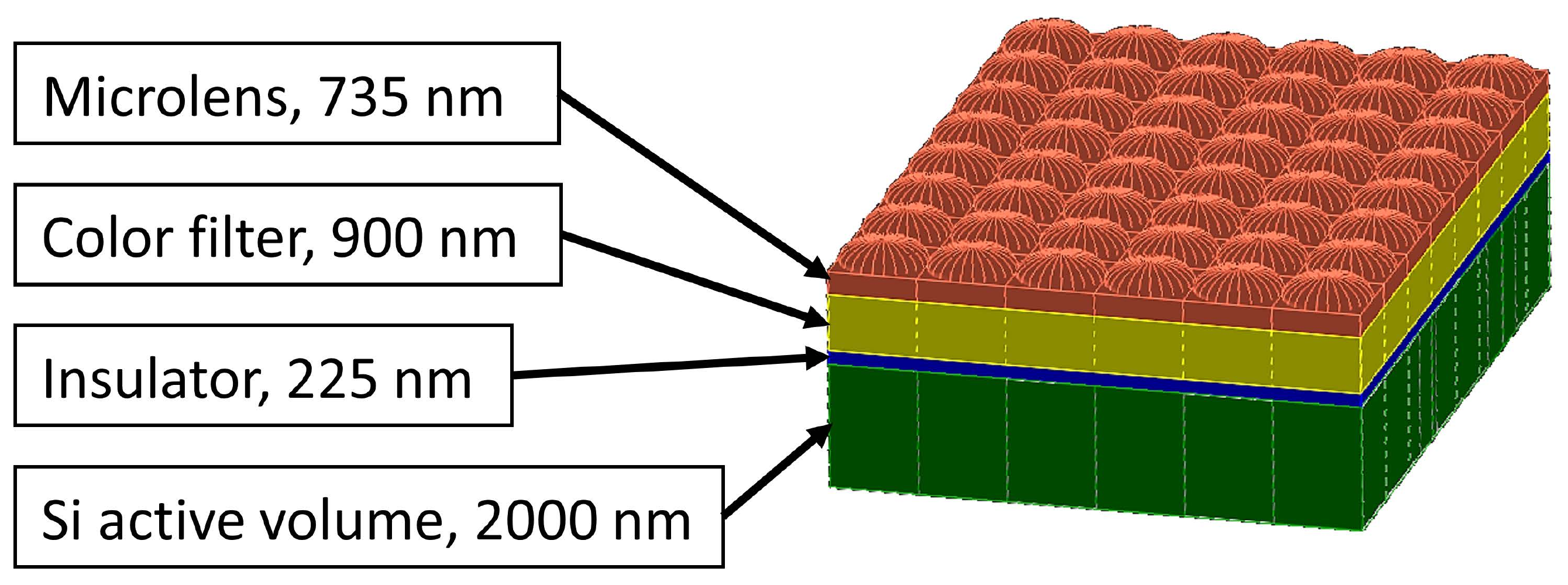}}
{\includegraphics[width=0.30\textwidth]{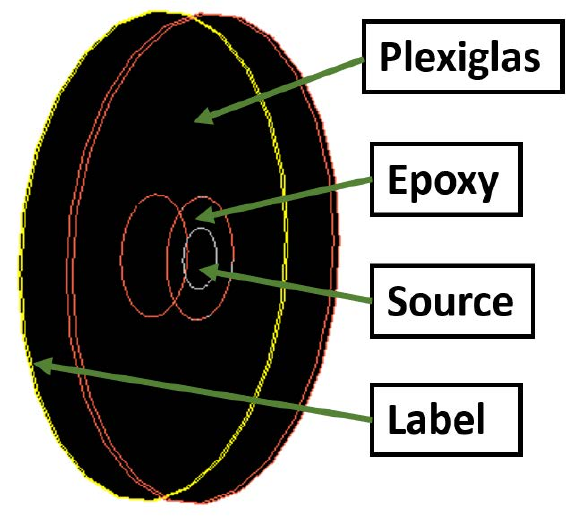}}
\caption{Simulated geometry of a fragment (6$\times$8 pixels) of the CMOS sensor (left) and the radioactive source (right). } 
\label{fig:geant4_geometry}
\end{figure}

A fragment of the simulated CMOS sensor geometry and the simulated radioactive source geometry are shown in figure \ref{fig:geant4_geometry}. The dark box was not simulated because it is large enough compared to the sensor, not affecting the measurement. Also, the distance between the source and the sensor is smaller than the dark box size.

The number of simulated events is equivalent to the number of disintegrations for a certain time. Given the activity of the radioactive sources defined in Sec. \ref{radio_source} the corresponding number of simulated events for the Sr90 source was 1514 and for the Cs137 source 3811 events, both equivalent to an exposure of 0.5 s.

\begin{figure}[htbp]
\centering
\includegraphics[width=0.5\textwidth]{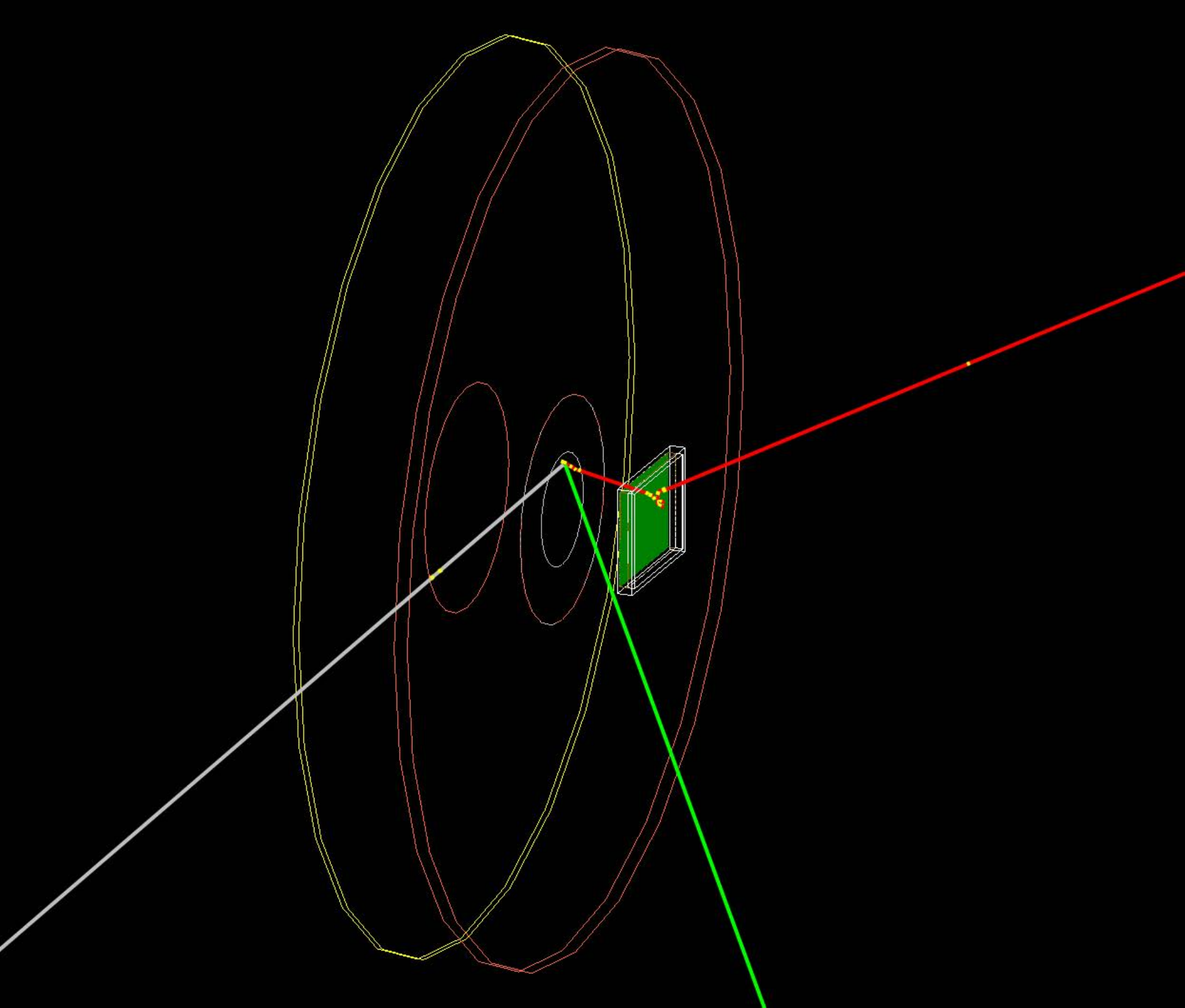}
\caption{Geant4 simulation of the particles emitted by a Cs137 source. Red lines represent electrons, the white lines are the electron antineutrino and the green lines are photons.} 
\label{fig:geant4_simu}
\end{figure}

Figure \ref{fig:geant4_simu} shows an example of 1 event of the Geant4 simulation of the particles emitted by a Cs137 radioactive source to the CMOS sensor, where the track left by the electron is red, the electron antineutrino track is white, and the photon is green.

A second step in our simulation, after Geant4 has given us the energy deposited in the pixel matrix is charge diffusion in the matrix, an inherent phenomenon in the photon-to-electron conversion process in CMOS sensors, such as the OV5647. 

This diffusion occurs when electrons generated by the interaction of particles from radiation (or light) are scattered before being collected by the sensor's pixels \cite{Esposito_2017, Everhart_1971, MCCARTHY1995538}. This scattering causes a loss of spatial resolution, as the charge generated by a single photon can be shared by several adjacent pixels, smoothing out the edges \cite{Esposito_2017, Everhart_1971, MCCARTHY1995538}. We have incorporated this phenomenon, modeled with a Gaussian distribution, into the Geant4 simulation \cite{Esposito_2017}.

When a photon or particle hits the semiconductor material of the sensor, a cloud of electrons and holes (electron-hole pairs) is generated due to the photoelectric effect or ionization \cite{Esposito_2017}. In the sensor region where there is no strong electric field (such as between the surface and the depletion region), known as the field-free region, charges can move by thermal diffusion before being collected due to interaction with the material \cite{Esposito_2017, Everhart_1971, MCCARTHY1995538}. 

We model the spatial diffusion of the deposited charge using a Gaussian function, where the number of affected pixels was limited to 9 in a 3 × 3 area, since it is a small effect. The central pixel receives most of the energy, while adjacent pixels contribute proportionally according to the Gaussian function:

\begin{equation} 
E_{\text{fraction}}(x, y) = \frac{E_{\text{dep}}}{2 \pi \sigma_{\text{total}}^2} \exp\left(-\frac{(x - x_0)^2 + (y - y_0)^2}{2 \sigma_{\text{total}}^2}\right)
\label{eq1}
\end{equation}

where:
\begin{itemize}
\item $E_{\text{fraction}}(x, y)$: is the fraction of energy or charge deposited in a pixel centered at $(x,y)$.

\item $E_{\text{dep}}$: is the total energy deposited in the current step.

\item $x_0$ and $y_0$: are the coordinates of the initial interaction point.

\item $\sigma_{\text{total}}$: is the total width of the Gaussian distribution and is calculated considering both the initial spread of the charge cloud and the diffusion in the field-free region \cite{Esposito_2017}.

\end{itemize}

The $\sigma_{\text{total}}$ is given by \cite{Everhart_1971, MCCARTHY1995538}:
\begin{equation} 
\sigma_{\text{total}} =\sqrt{{\sigma_{\text{i}}}^2 + 
 {\sigma_{\text{ff}}}^2}
\label{eq2}
\end{equation}

where $\sigma_{\text{i}}$ is the initial radius of the charge cloud and $\sigma_{\text{ff}}$ is the diffusion in the field-free region. 

The initial radius of the charge cloud is calculated by \cite{Esposito_2017, Everhart_1971, MCCARTHY1995538}: 
\begin{equation} 
\sigma_i = k \cdot E_{\text{dep}}^{\alpha}
\label{eq3}
\end{equation}

where $k$ and $\alpha$ are empirical constants used to convert energy into a spatial range, these constants depend on factors such as the density of the material, its crystalline structure, and the properties of the medium in which the electrons move \cite{Everhart_1971, MCCARTHY1995538}.

The constant $k$ is related to the initial diffusion of the charge. This constant is derived from experimental studies and is specific to the material as well as the conditions under which the experiment was performed \cite{Everhart_1971, MCCARTHY1995538}. 

The constant $\alpha$ is an empirical exponent that describes how the range of electrons changes with respect to energy. In most materials, this value lies between 1 and 2, depending on the electron energy and the characteristics of the material \cite{Everhart_1971, MCCARTHY1995538}.

The diffusion in the field-free region is calculated by \cite{Everhart_1971, MCCARTHY1995538}:
\begin{equation} 
\sigma_{ff} = \frac{z_{ff}}{2} \sqrt{1 - \left( \frac{z_0 - z_{ff}}{z_{ff}} \right)^2}
\label{eq4}
\end{equation}

where $z_{ff}$ is the thickness of the field-free region, and $z_0$ is the z-coordinate of the interaction point. 

The Gaussian diffusion parameters (i.e. $z_{ff}$, $k$, and $\alpha$) of the OV5647 CMOS sensor are not well known. To determine their most probable values, we perform a minimization search within ranges suggested by the literature and constrained by geometrical limits, comparing the simulation results with experimental data. However, since charge diffusion has only a marginal effect on the overall simulation, this minimization does not inherently guarantee agreement with the experimental data. Instead, the main source of agreement arises from the robust validation of the Geant4 simulation. 

Next we define the parameters ranges. Given a total sensor thickness ($z_{max}$) of 2 $\mu$m, the field-free region ($z_{ff}$) can be between 0.2 $\mu$m and 1 $\mu$m (10-50\% of $z_{max}$). Due to the absence of semiconductor doping and material properties, the empirical constants $k$ and $\alpha$ are also estimated, which were varied between 0.002 to 0.062 for $k$ and 1.73 to 1.77 for $ \alpha $. Simulations covering the parameters grid are run with different values of $z_{ff}$, $k$, and $\alpha$, ensuring that the sensor dimensions are not exceeded. The grid steps for these parameters were: 0.2 for $z_{ff}$, 0.012 for $k$, and 0.005 for $\alpha$. Optimal values for these three parameters are then found by minimizing the difference between simulation and experimental measurements. In our case, the best-fit values for the OV5647 are $z_{ff}$ = 1 $\mu$m, $k$ = 0.002 $\mu$m/keV, and $\alpha$ = 1.75. The $\alpha$ value coincides with the one found for silicon in \cite{Esposito_2017, MCCARTHY1995538}.

After applying the charge diffusion algorithm, we convert the remaining deposited energy into electron-hole pairs in Si, using the factor of 3.6 eV (see Sec. \ref{detect_cmos}) per pair. From the number of pairs, we extrapolate the number of electrons used in the rest of the simulation chain.  
After the deposited energy per pixel is transformed into a number of electrons, these electrons are converted to digital values by an Analog-to-Digital Converter (ADC) using the sensor's WC value, which limits the maximum number of electrons a pixel can store. As we have pointed out, this number is 4300 electrons, corresponding to the WC, achieved with an analog gain of 8 which is the maximum value that the OV5647 sensor can reach to ensure stable data acquisition. Gains greater than 8 could lead to ADC saturation, as indicated in the datasheet \cite{datasheets_rasp_camera_WC, FWC_OV5647}.

To calculate the minimum number of electrons for an ADC count, we divide the total FWC (4500 electrons) by the usable range of digital values of the ADC, which in our case is 10-bit, equivalent to a range of 1023 different values ($2^{10} - 1$), resulting in approximately 4.4 electrons per ADC count. However, to maintain the integrity and precision of the conversion process, this value is rounded to the next integer, obtaining in this case a minimum of 5 electrons per ADC count.

\section{Data analysis} \label{dat_analysis}

The electronics of the camera produce a non-negligible noise during the pixel readout, especially when working with low signals. This noise causes, in our case, that 99.48 \% of the pixels have a signal detected in the experimental data. 

To reduce this background noise in the experimental data a two-step procedure was applied: (1) the average ADC for each pixel from the background data (11.5 ADC on average) was subtracted from the experimental measurements, reducing it to 43.6\% of active pixels. (2) After the subtraction, a five standard deviations ADC threshold for each pixel from the background (5$\sigma$=21.1 ADC on average) was applied, reducing it to 0.0047\% of active pixels (i.e. 236474 active pixels). From now on, the aforementioned process, that we call as Standard Noise Reduction (SNR), will be applied by default to all the experimental data presented in our work, unless otherwise specified.

Assuming that this background noise is stable and evenly distributed in the pixel matrix, this procedure should reduce most noise. This method does not affect signal events with large ADC values. However, this is not the case for low ADC signals. Nevertheless, it is not possible to distinguish between one pixel with a low ADC coming from background from a similar one from signal. 

Using the OpenCV (Open Computer Vision) libraries \cite{opencv}, clusters (i.e., a pixel or groups of pixels with a signal different than zero and adjacent to each other) are searched for in all frames. In our analysis, we will study three observables: the ADC distribution, the cluster size, and the maximum ADC per cluster.

When analyzing the cluster size (defined by its number of pixels), we found that the background noise is mostly present in the smallest clusters of 5 pixels at most, and we cannot differentiate between the signal and the background for clusters of similar size. Therefore, on top of the SNR procedure, we make a cut on the cluster size discarding clusters composed by less than or equal to 5 pixels, removing 100\% of the remaining background clusters. Using this filtered sample, we analyze the maximum ADC per cluster. The application of this cut will enable us to compare this reduced data sample, nearly devoid of background clusters, with the Geant4-simulated signal that also underwent the 5-pixel cluster cut. 

The result of our statistical analysis will be shown in terms of the reduced $\chi^2_\nu$ defined as $\chi^2/\nu$ where $\nu$ is the number of degrees of freedom.  

\section{Results} \label{resul}

We present the results of the comparison between our experimental data and the Geant4-simulation of the camera when it is irradiated by a radioactive source. As we already mentioned, this analysis will be performed for three observables: the ADC distribution, the cluster size, and the maximum ADC per cluster, considering as sources: Sr90 and Cs137. Furthermore, by moving the radioactive sources to various distances, we will investigate the square inverse distance law of irradiation. 

In general, the experimental data show the statistical uncertainty, while the simulation uncertainty propagates the systematic source activity uncertainty, assumed to be a 20\% error according to the datasheet \cite{spectrumtechniques}, combined with the standard deviation from ten simulations. As mentioned before, the experimental data shown in all the plots has already been filtered by our SNR procedure. 

\begin{figure}[htbp]
\centering
{\includegraphics[width=0.47\textwidth]{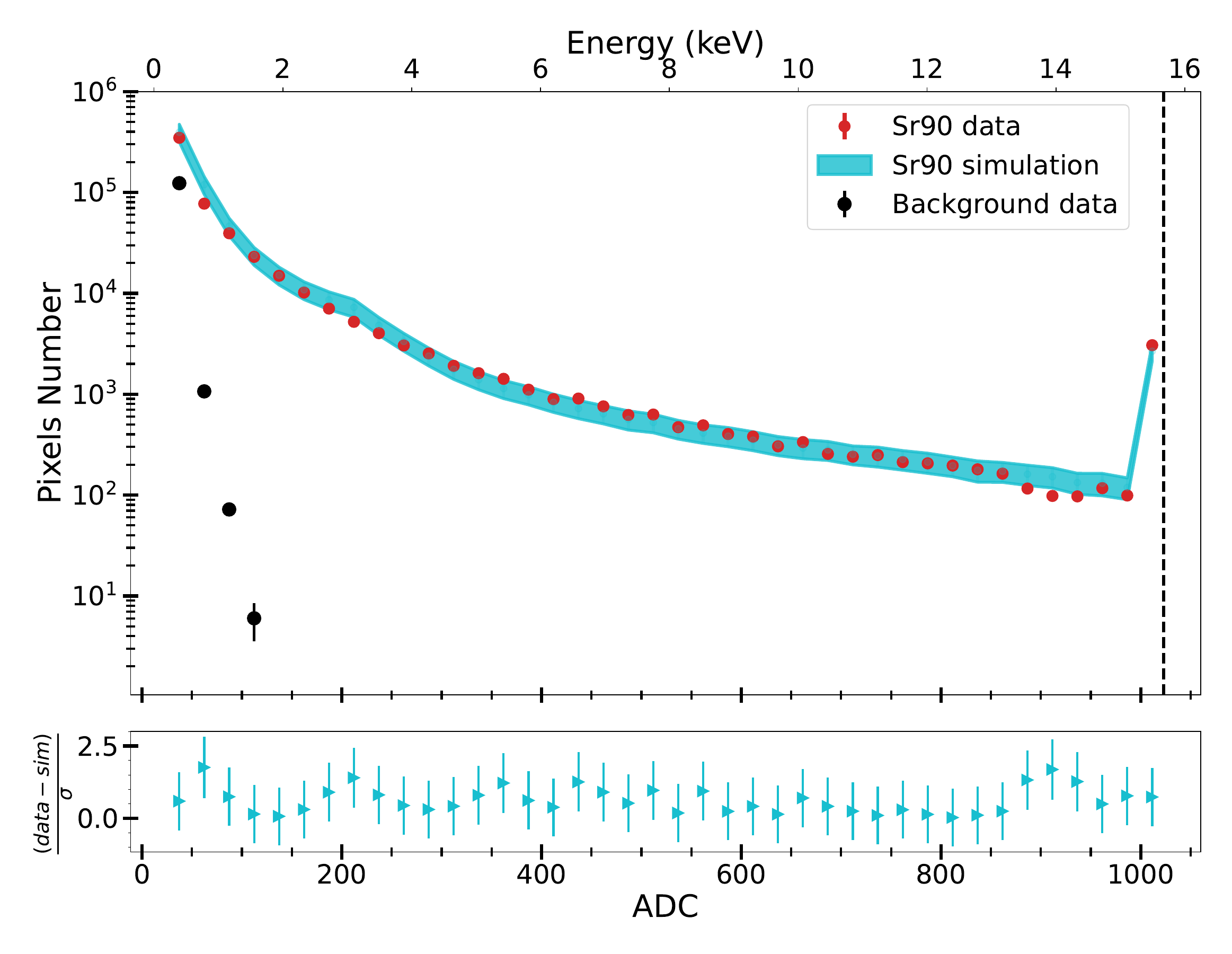}}
{\includegraphics[width=0.47\textwidth]{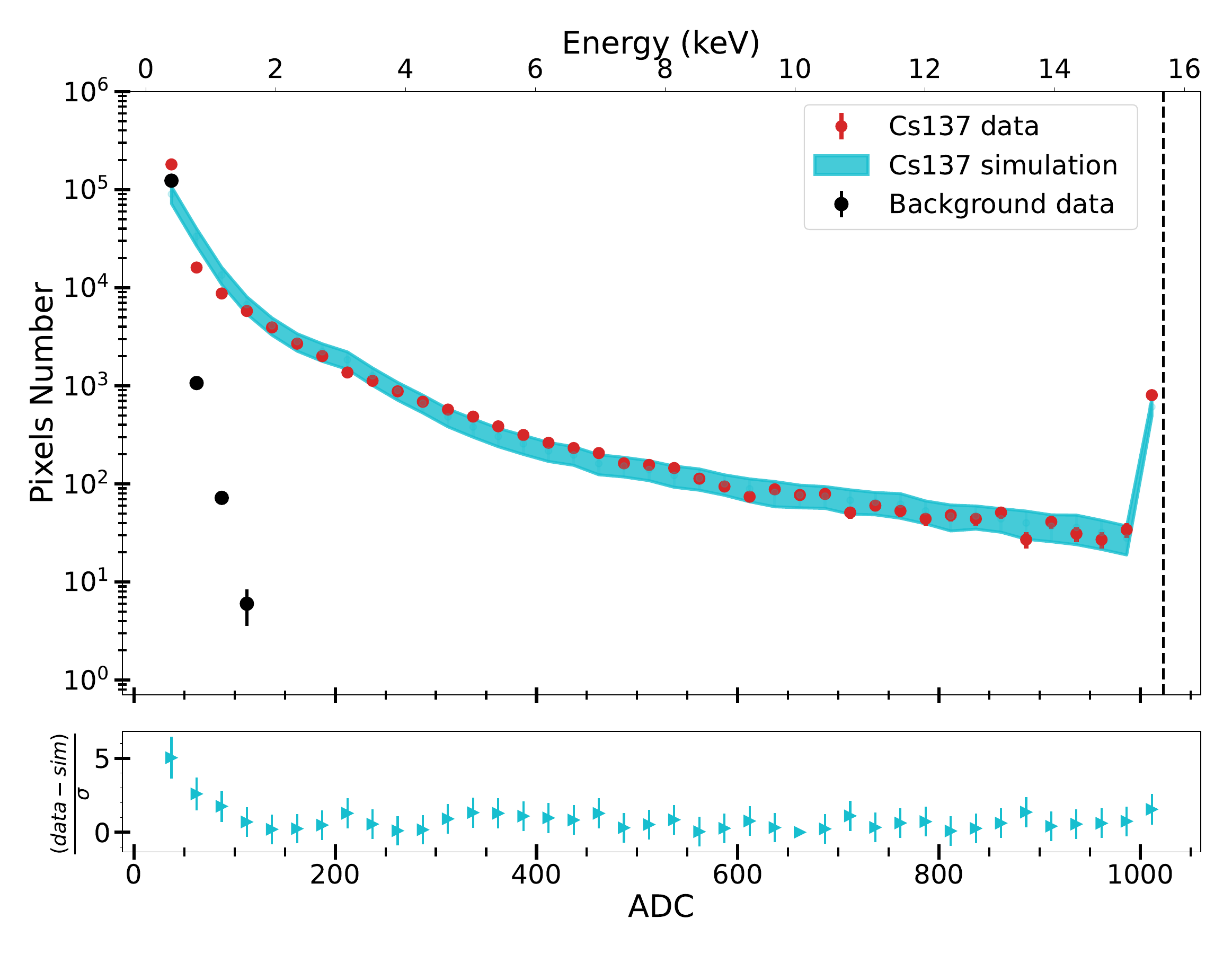}}
{\includegraphics[width=0.47\textwidth]{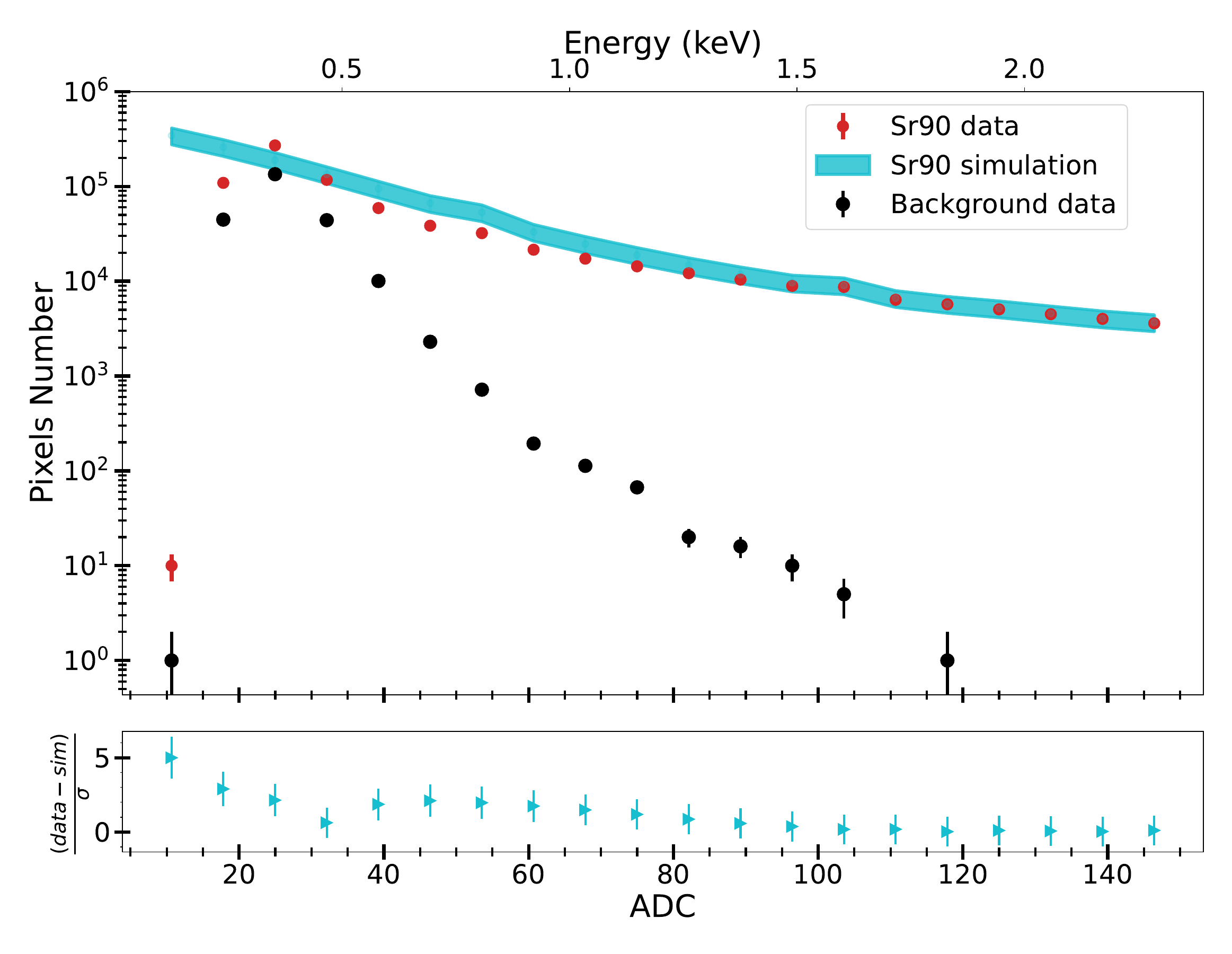}}
{\includegraphics[width=0.47\textwidth]{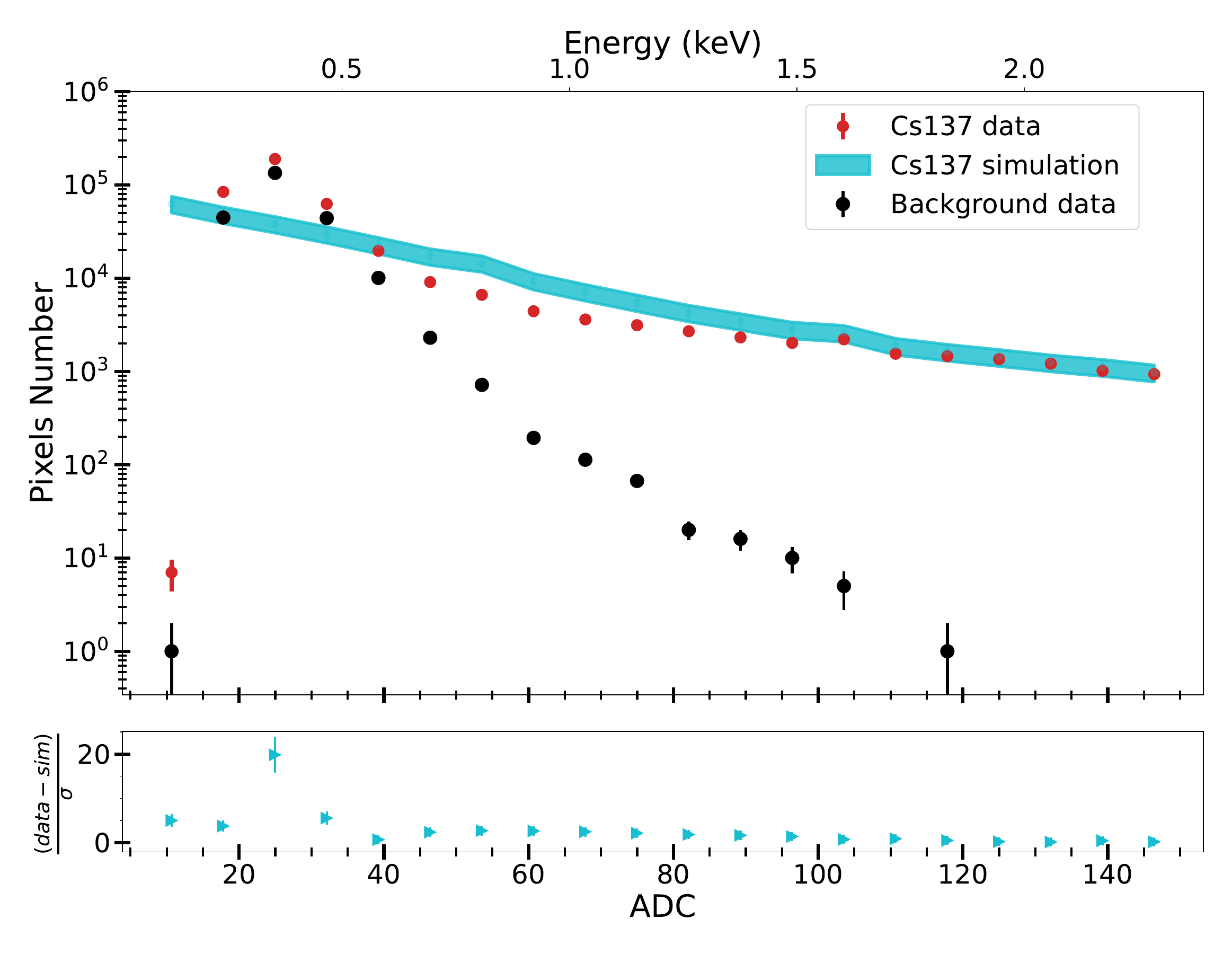}}
\caption{
ADC number distribution at 0 mm between sensor and source, for 500 s exposure time for the experimental and simulated data for Sr90 (top left) and Cs 137 (top right), the experimental data applying the SNR procedure and comparing with pure simulation, the vertical dotted line shows the saturation level of the CMOS sensor at 1023 ADC equivalent to 15.48 keV. At the bottom a zoom of the same distributions for the region between 1 and 150 ADC for Sr90 (bottom left) and Cs137 (bottom right). 
}
\label{fig:plot_hist_ADC}
\end{figure}

Figure \ref{fig:plot_hist_ADC}, compares the ADC distributions at the pixel level for all cluster sizes for the experimental data and its corresponding simulation, for both Sr90 and Cs137. It is important to note that the signal saturation is reached at 1023 ADC, equivalent to 15.48 keV of deposited energy. The background noise only attains up to 115 ADC. 

In the plots at the top, the experimental data is presented after applying the SNR procedure. In both plots (left and right), the background is noticeable and located at the first bins. In the case of Sr90, the average number of active pixels for the experimental data, after applying the SNR procedure, is 0.011\%, while for the Geant4 signal simulation, it is 0.013\%. Analogously, for Cs137, the average number of active pixels for the experimental data is 0.005\%, and for the simulation, 0.003\%, which is a lower number than for Sr90, even if the activity of the Cs137 source is higher. 

The electrons emitted by Sr90 have a higher maximum energy than those from Cs137, resulting in a greater number of detections with the Sr90 source than with the Cs137 source, since the higher energy electrons have a higher probability of passing through the different layers of the material until arriving to the detector. In comparison, Cs137 electrons, having lower energy, have a lower probability of depositing energy in the detector \cite{nncs, knoll}. The latter is clearly illustrated in the plots at the bottom of Figs. \ref{fig:plot_hist_ADC}, in which we are zooming in the region up to 150 ADC, where the ADC distribution for Sr90 is higher than for Cs137.

To remove the background at the first bins after applying the SNR procedure, as shown in Figs \ref{fig:plot_hist_ADC}, we cut the events below 100 ADC, since we cannot tell apart signal from background. After this cut, the background left is minimal, and the Sr90 average number of active pixels is reduced to 0.002\%, while for Cs137 reaches 0.0005\%, getting for both radioactive sources an excellent agreement in terms of the number of active pixels between the experimental data and the simulation.

\begin{figure}[htbp]
\centering
{\includegraphics[width=0.47\textwidth]{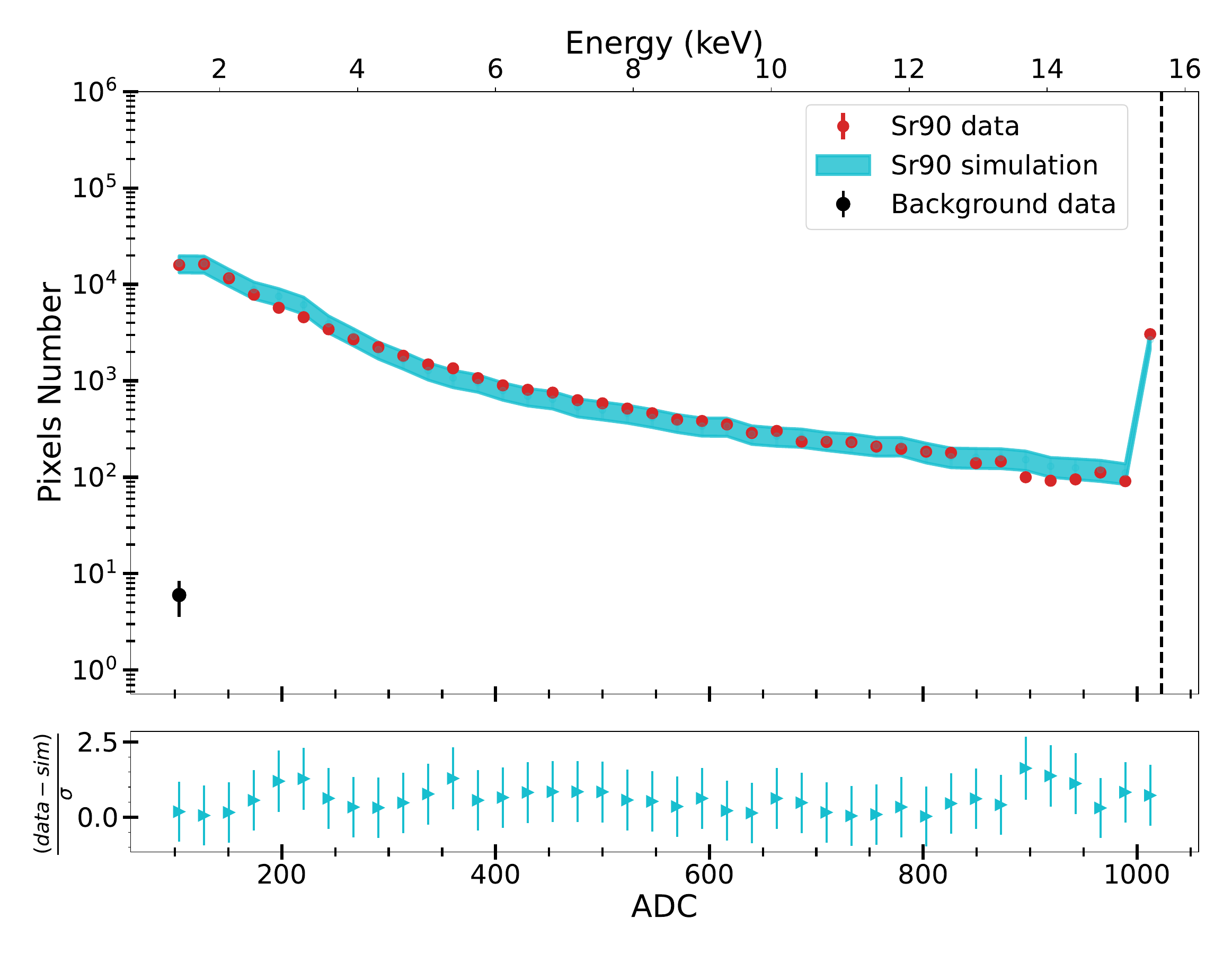}}
{\includegraphics[width=0.47\textwidth]{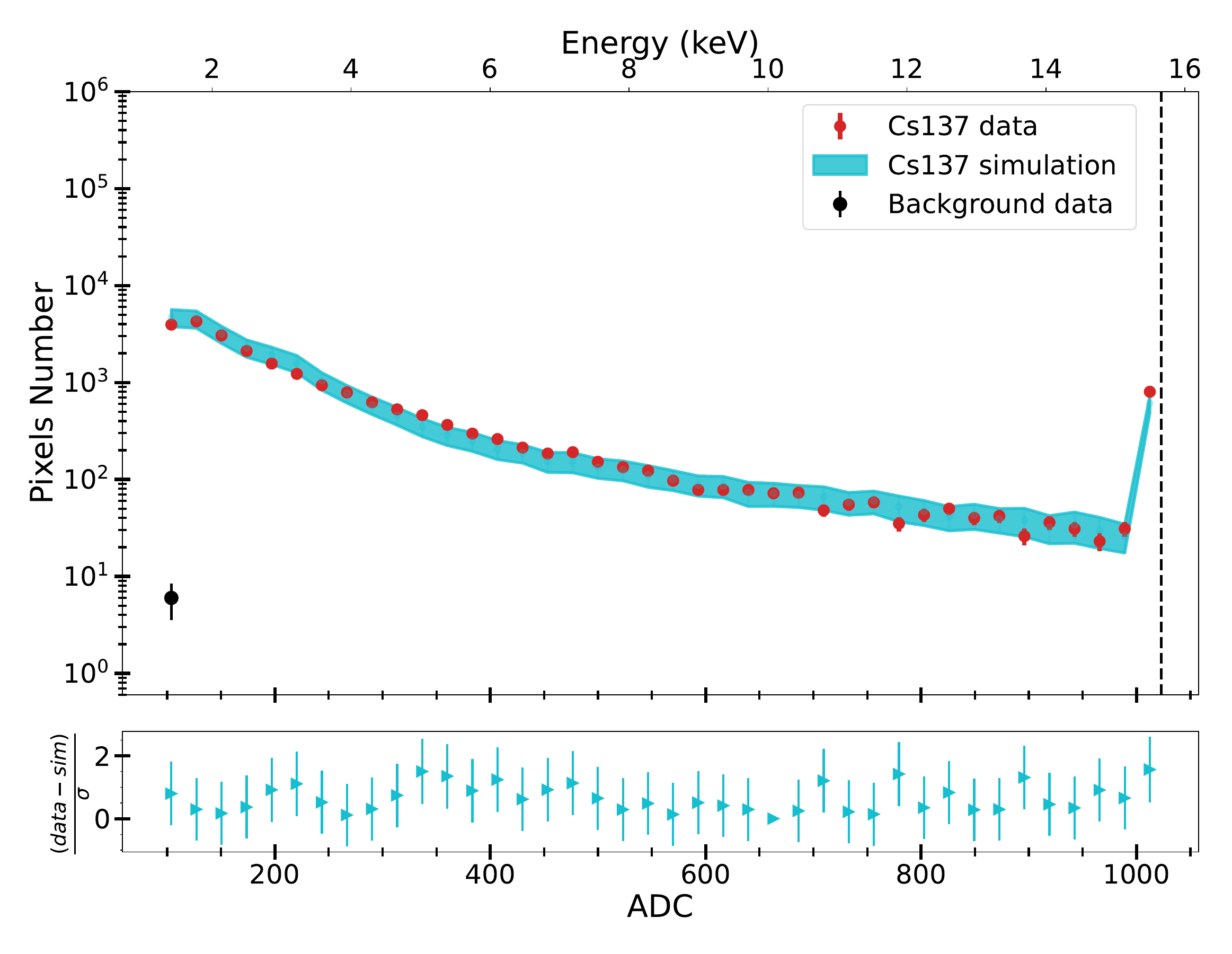}}
\caption{
ADC number distribution for the experimental and simulated data with a cut to remove values less than or equal to 100 ADC for Sr90 (left) and Cs 137 (right), the experimental data applying the SNR procedure and comparing with pure simulation at 0 mm between sensor and source, for 500 s exposure time. The vertical dotted line shows the saturation level of the CMOS sensor at 1023 ADC equivalent to 15.48 keV.
}
\label{fig:plot_hist_ADC_cut}
\end{figure}

In figure \ref{fig:plot_hist_ADC_cut} we display the ADC distribution after applying the aforementioned cut. Here, we can see a very good agreement between the ADC distributions from the experimental data and the simulation for both sources. This is confirmed by the reduced chi-square test ($\chi^2_\nu$) for these ADC distributions, where we obtain a $\chi^2_\nu$ for Sr90 of 0.45, while for Cs137 is 0.48.

\begin{figure}[htbp]
\centering
\includegraphics[width=0.7\textwidth]{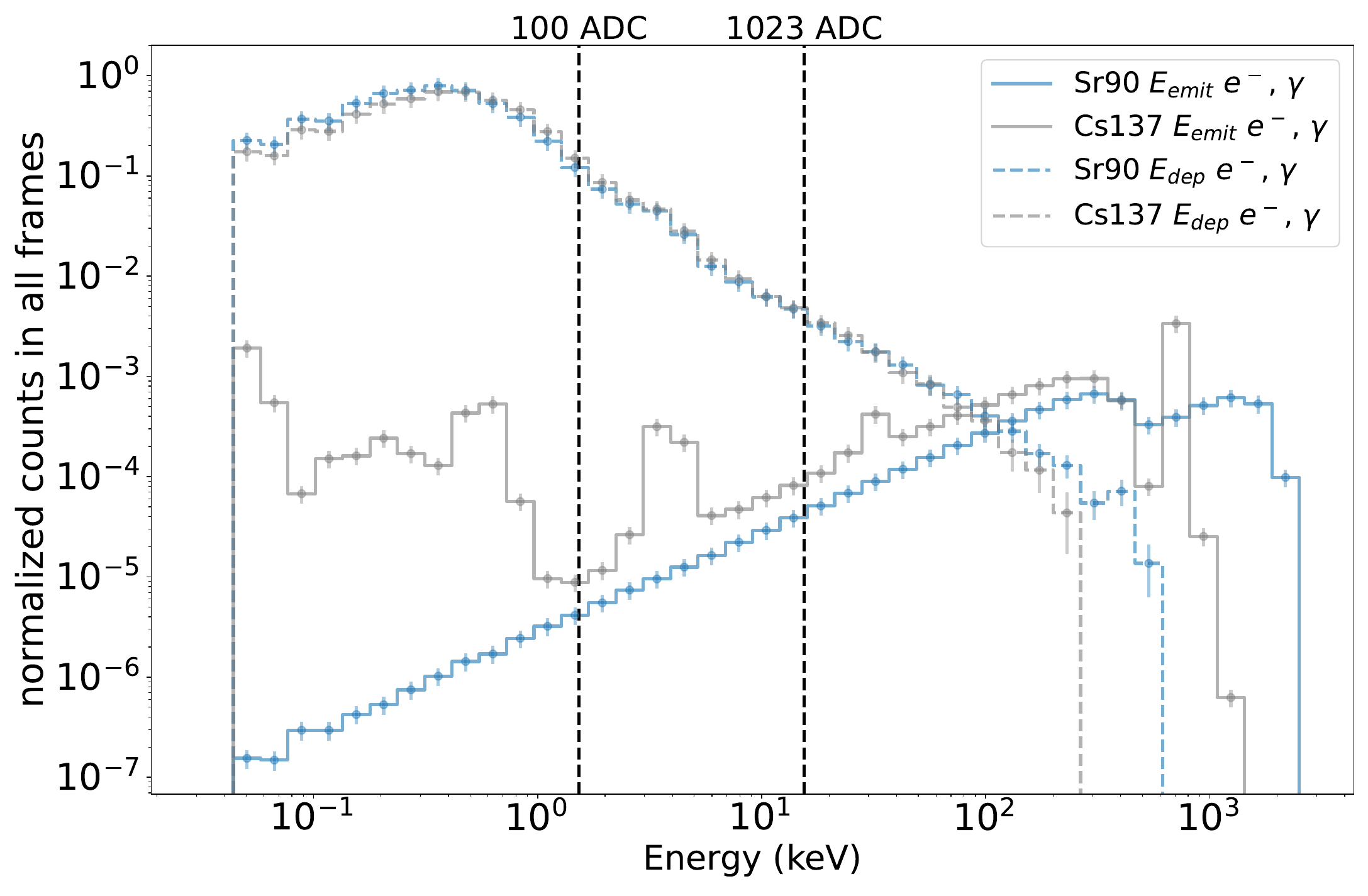}
\caption{
Normalized energy distributions, without cuts, of electrons and photons emitted from Sr90 and Cs137 and their corresponding deposited energy in the sensor pixel matrix, from the Geant4 simulation. The vertical dashed lines limit the sensor sensitivity, showing the low ADC cut that removes background and the upper ADC sensor saturation. 
}
\label{fig:emitted_deposited}
\end{figure}

To analyze the effect of the ADC cut and saturation on the signal, we analyze its dynamic range from simulation, shown in figure \ref{fig:emitted_deposited}. The energy deposited by electrons and photons emitted from Sr90 and Cs137 decays covers a much larger spectrum than the range detectable with the sensor (from the lower 100 ADC cut (1.5 keV) to the maximum 1023 ADC (15.48 keV)). In the sensitive range of the sensor the normalized distributions of both radioactive sources have the same shape. Only at the highest energies there is a difference. However that area is not reachable for the sensor, since it lies beyond the saturation peak. The lower energy depositions below the 100 ADC cut follow the same trend for Sr90 and Cs137. Thus, even if we are cutting a large part of the low energy signal where the background overlaps, we are not losing differentiation capabilities. In contrast, the energy distributions from the initial emitted electrons and photons do have a different shape, however, it is not translated into the deposited energy. 

\begin{figure}[htbp]
\centering
{\includegraphics[width=0.47\textwidth]{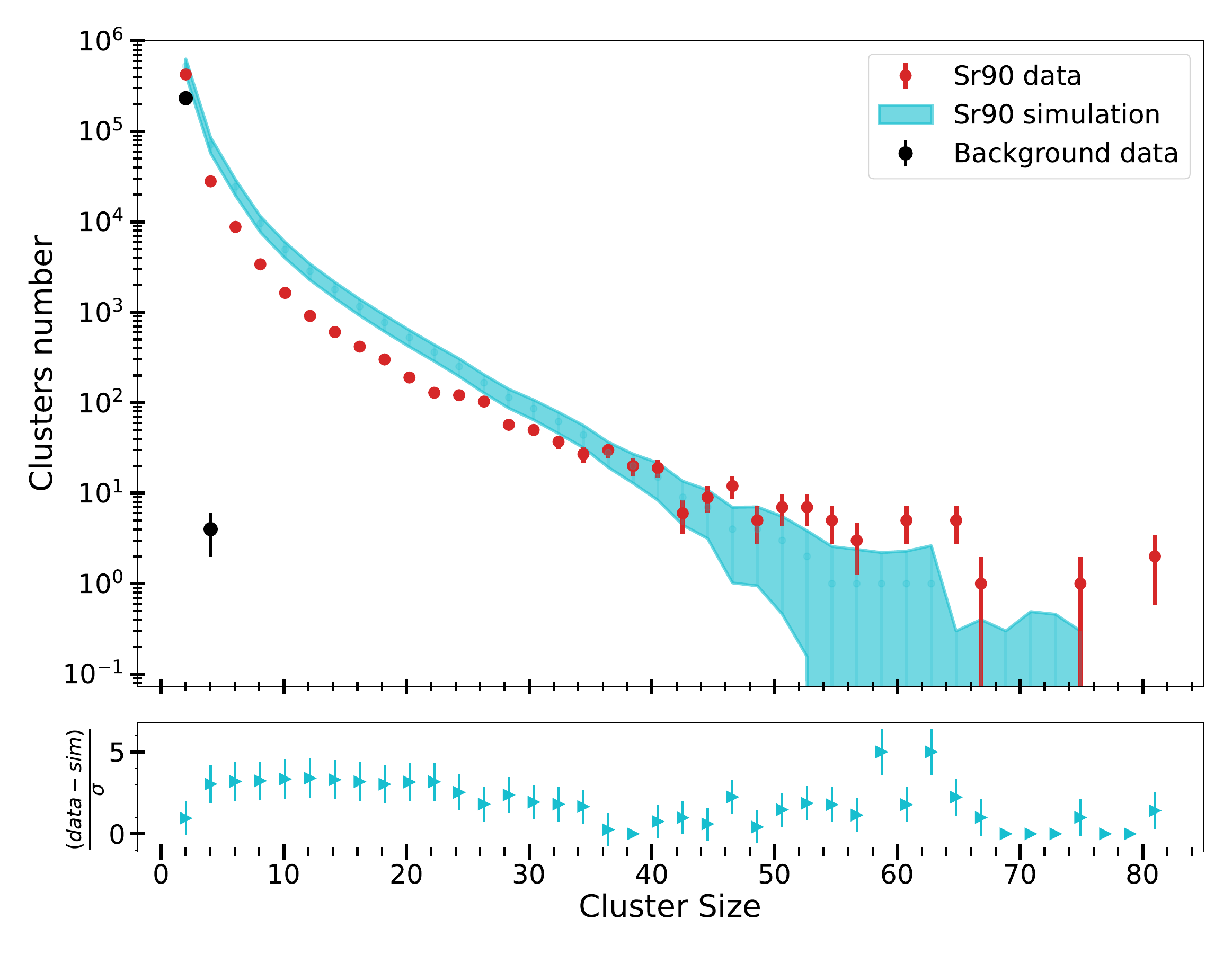}}
{\includegraphics[width=0.47\textwidth]{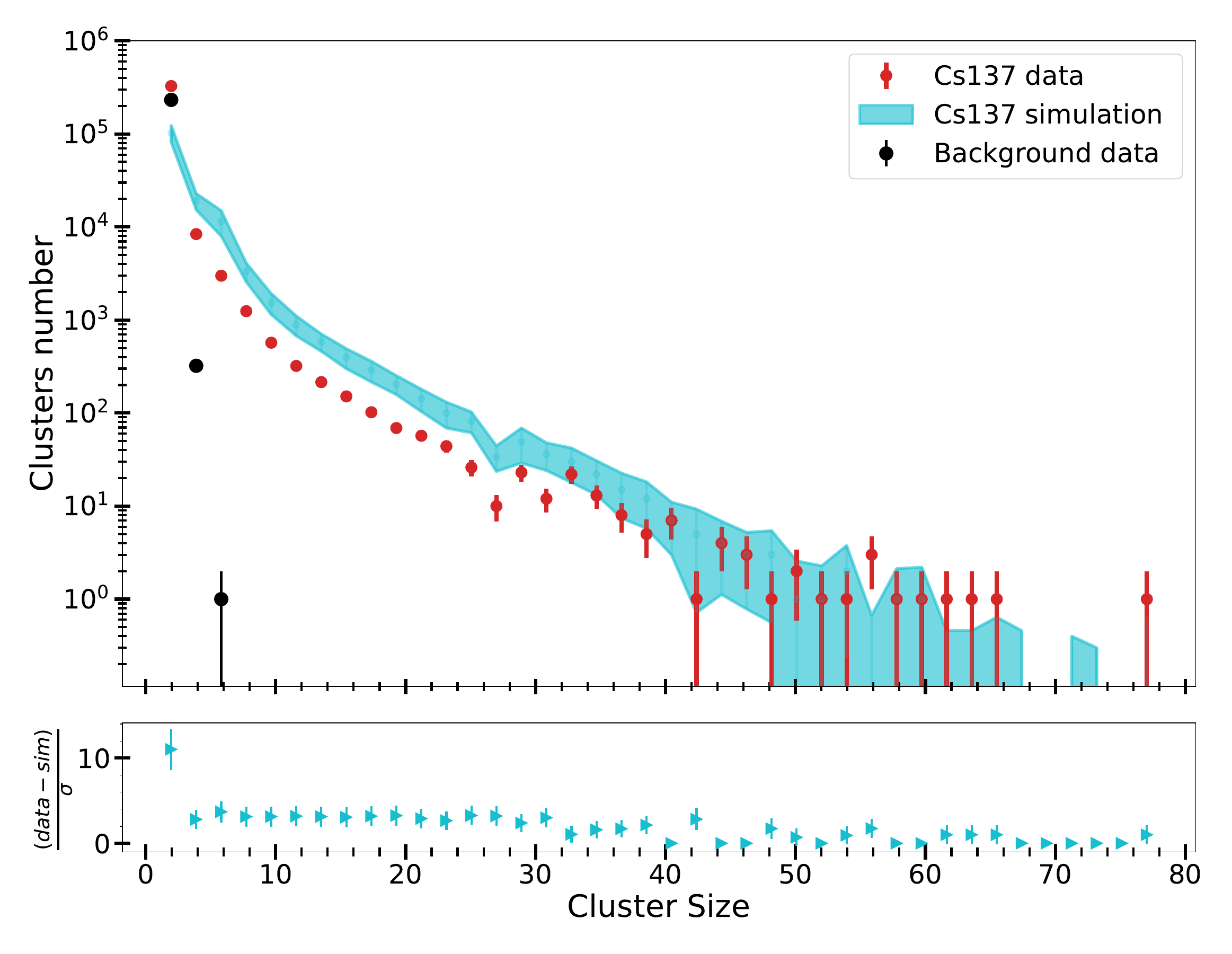}}
{\includegraphics[width=0.47\textwidth]{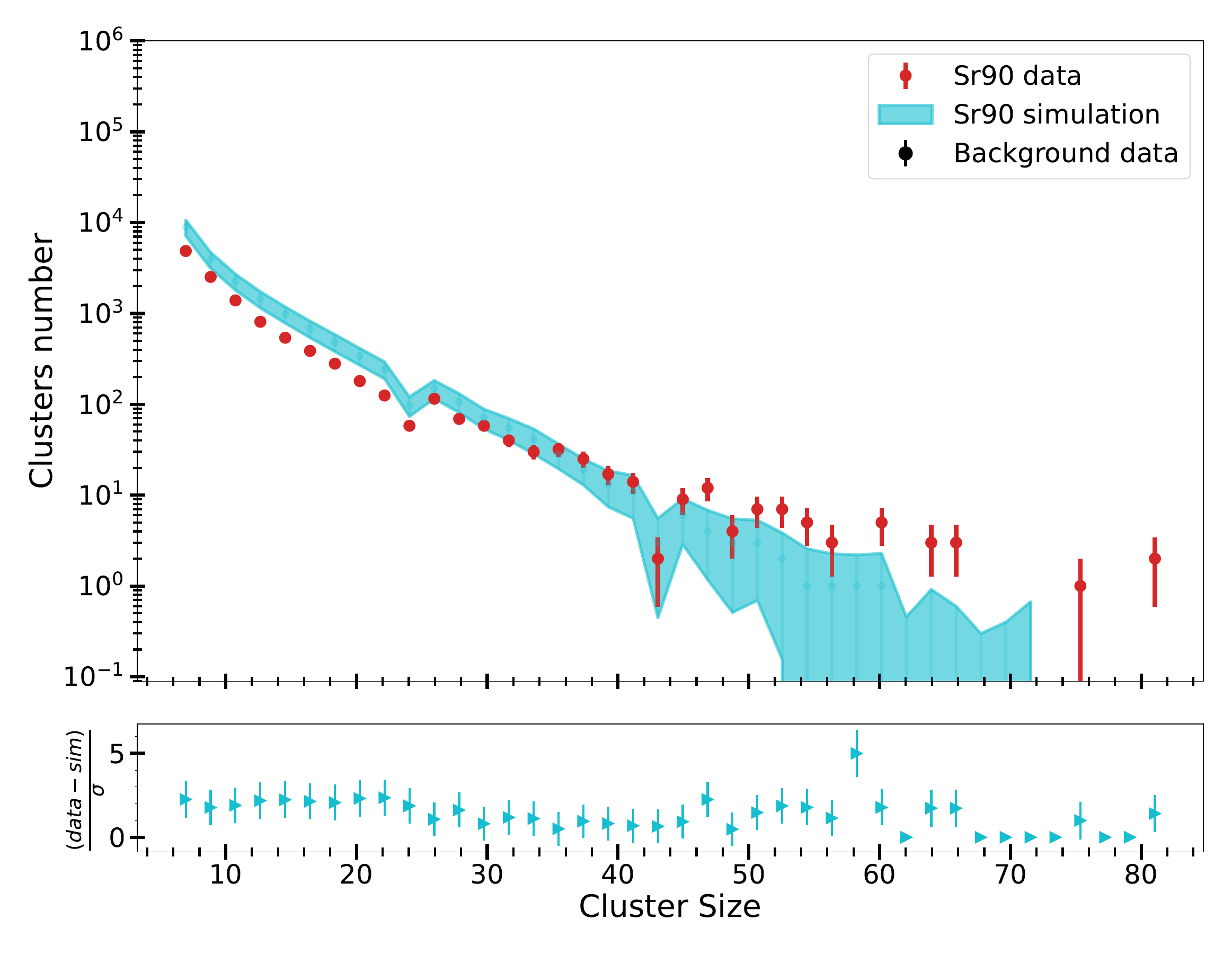}}
{\includegraphics[width=0.47\textwidth]{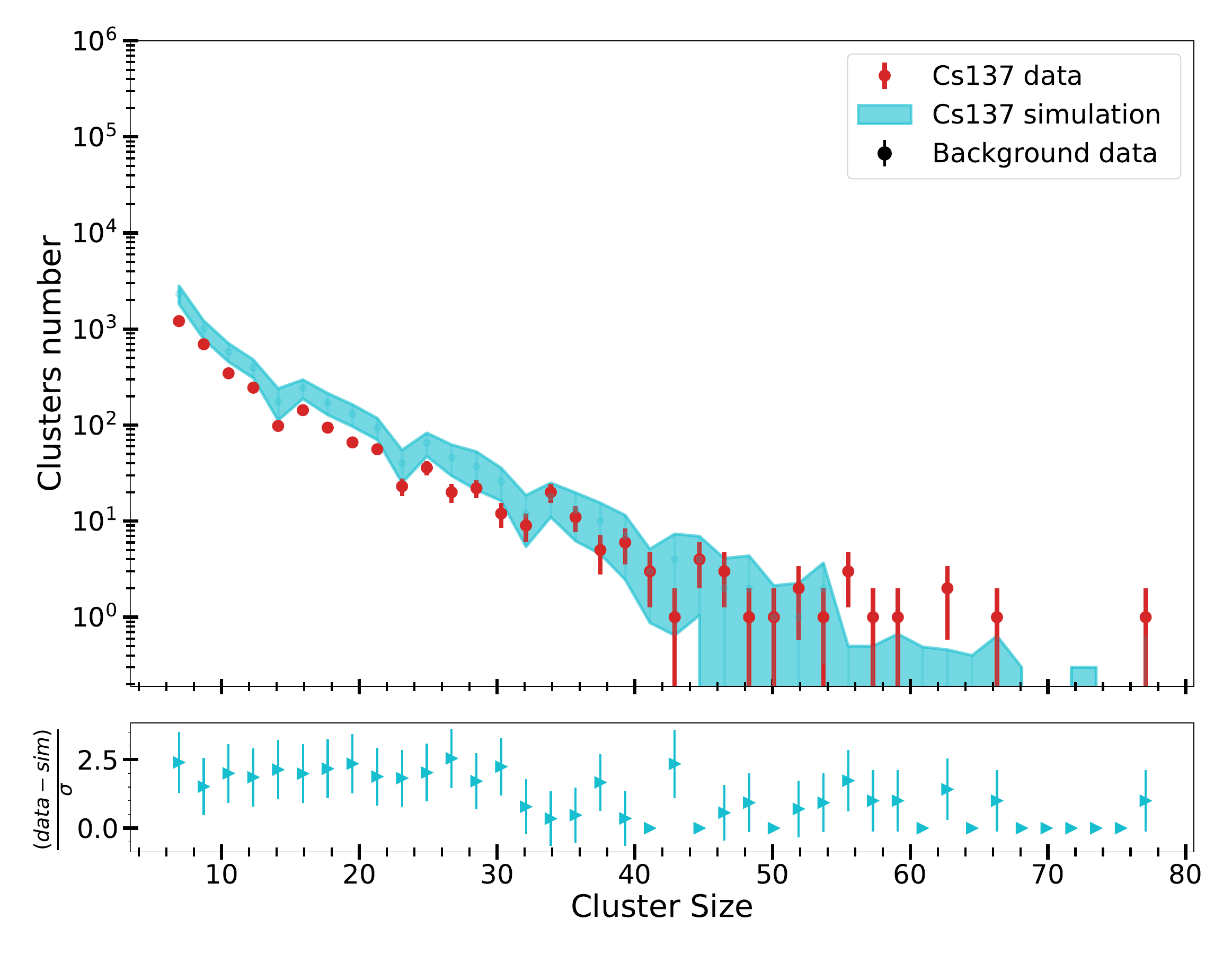}}
\caption{
Cluster size distribution for the experimental and simulated data for Sr90 (top left) and Cs 137 (top right), the experimental data applying the SNR procedure and comparing with pure simulation. At the bottom are the same distributions applying the CC procedure for Sr90 (bottom left) and Cs 137 (bottom right) at 0mm between sensor and source for 500 s exposure time.
}
\label{fig:plot_hist_Cluster_size}
\end{figure}

The top panels in figure \ref{fig:plot_hist_Cluster_size} show the cluster size distribution for experimental data and background after the SNR procedure is applied for both sources. The cluster size distribution shows that the background is present up to the five-pixel cluster, where most of them are located below the five-pixel cluster size. Therefore, on top of the SNR procedure, we apply two cuts: one is for removing clusters with less or equal to a five-pixel cluster. The other cut eliminates clusters where one or more of its pixels is less than or equal to 100 ADC. We will call the combination of both cuts Cluster Cuts (CC). The 100 ADC cut will also remove part of the signal and modify the structure and total deposited energy of the original clusters. However, since the sensor has no energy resolution capabilities, as explained later, the drawback of this cut is reduced.

The application of all these cuts is shown in the bottom panels of figure \ref{fig:plot_hist_Cluster_size}. Here, we observe the improvement of the agreement between the cluster size distribution for data and simulation after the cuts are applied. Considering these distributions, the $\chi^2_\nu$ value for Sr90 is 1.77, and for Cs137 is 1.16. These values indicate that there is a slight discrepancy between the experimental data and the simulation.     

\begin{figure}[htbp]
\centering
{\includegraphics[width=0.47\textwidth]{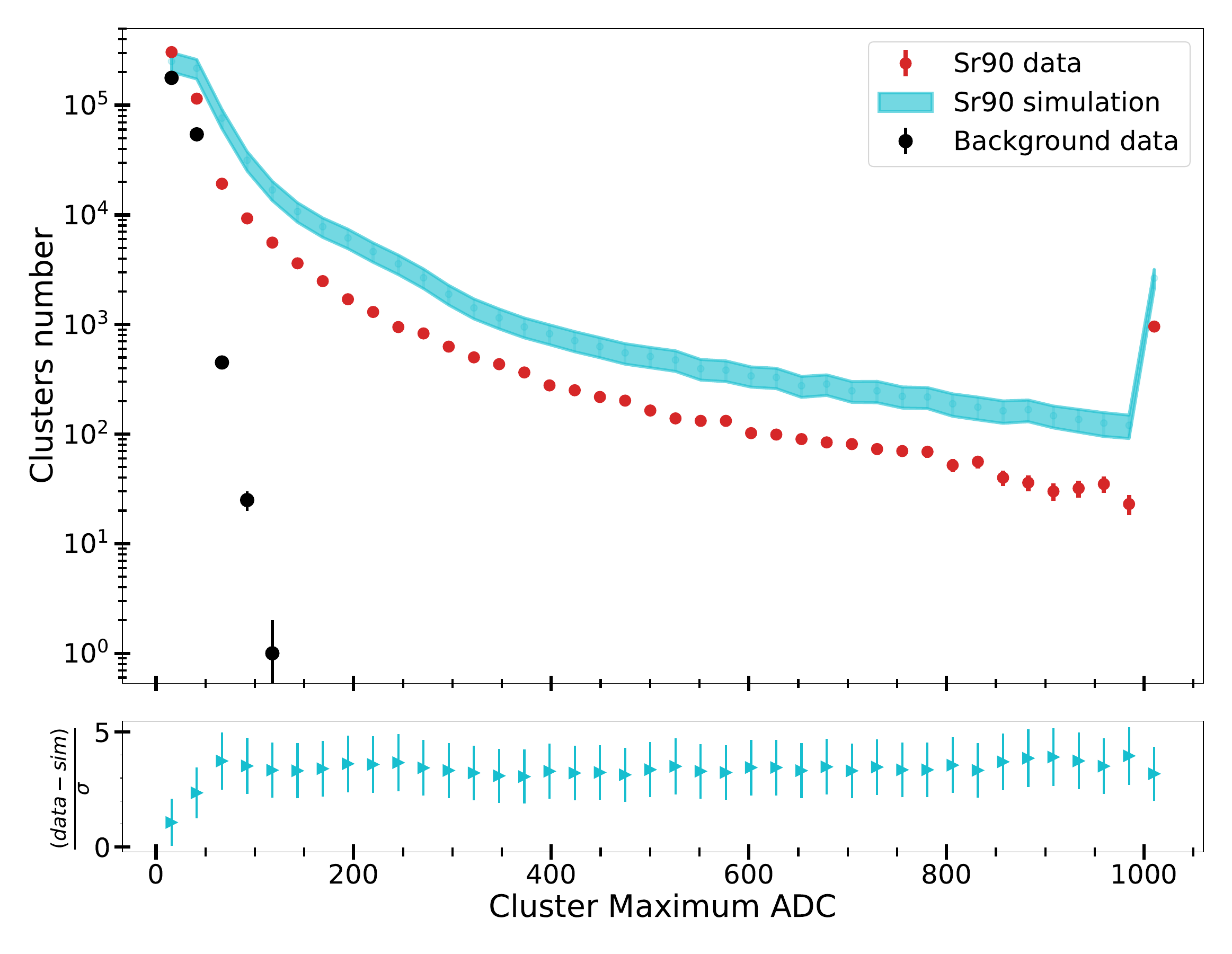}}
{\includegraphics[width=0.47\textwidth]{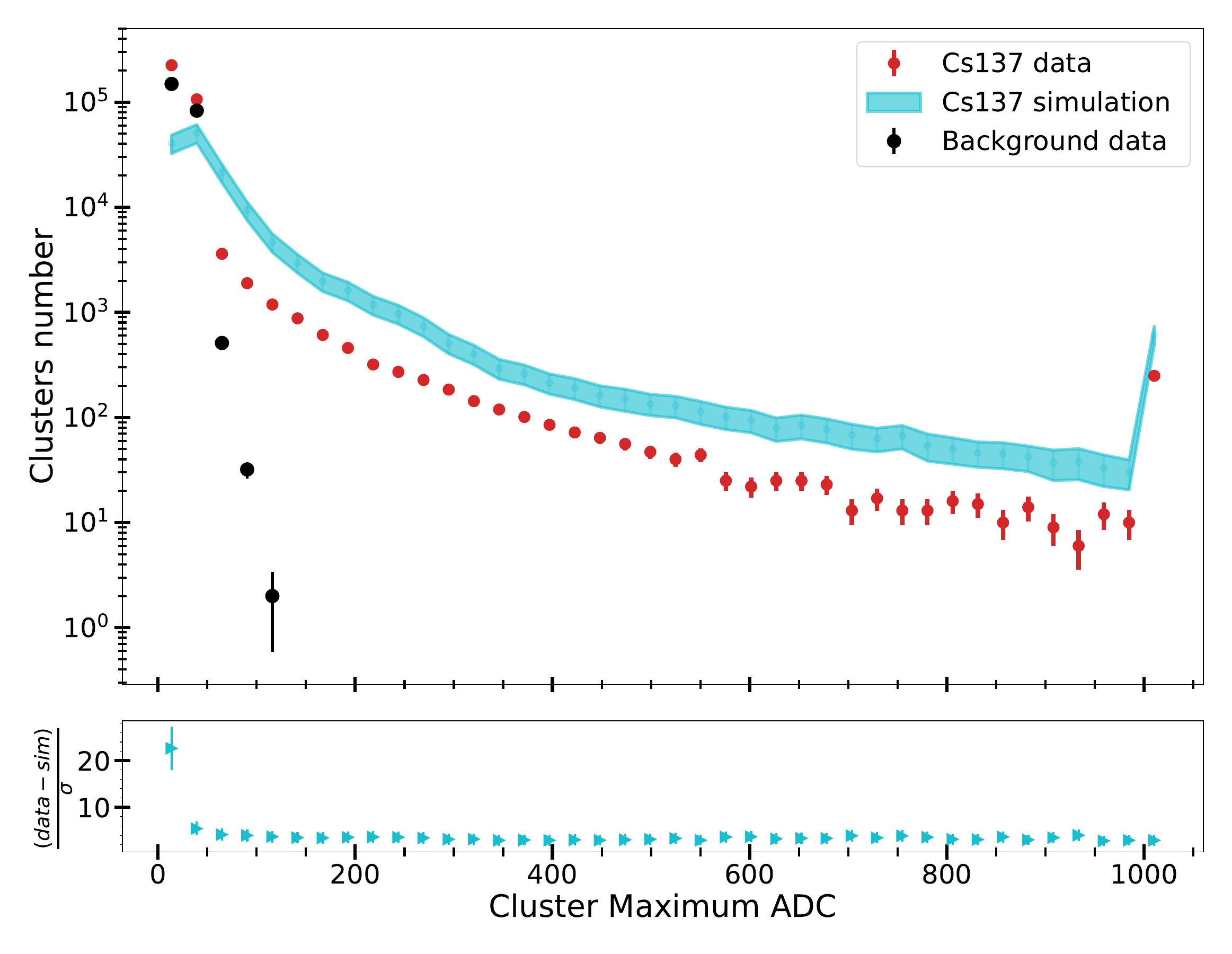}}
{\includegraphics[width=0.47\textwidth]{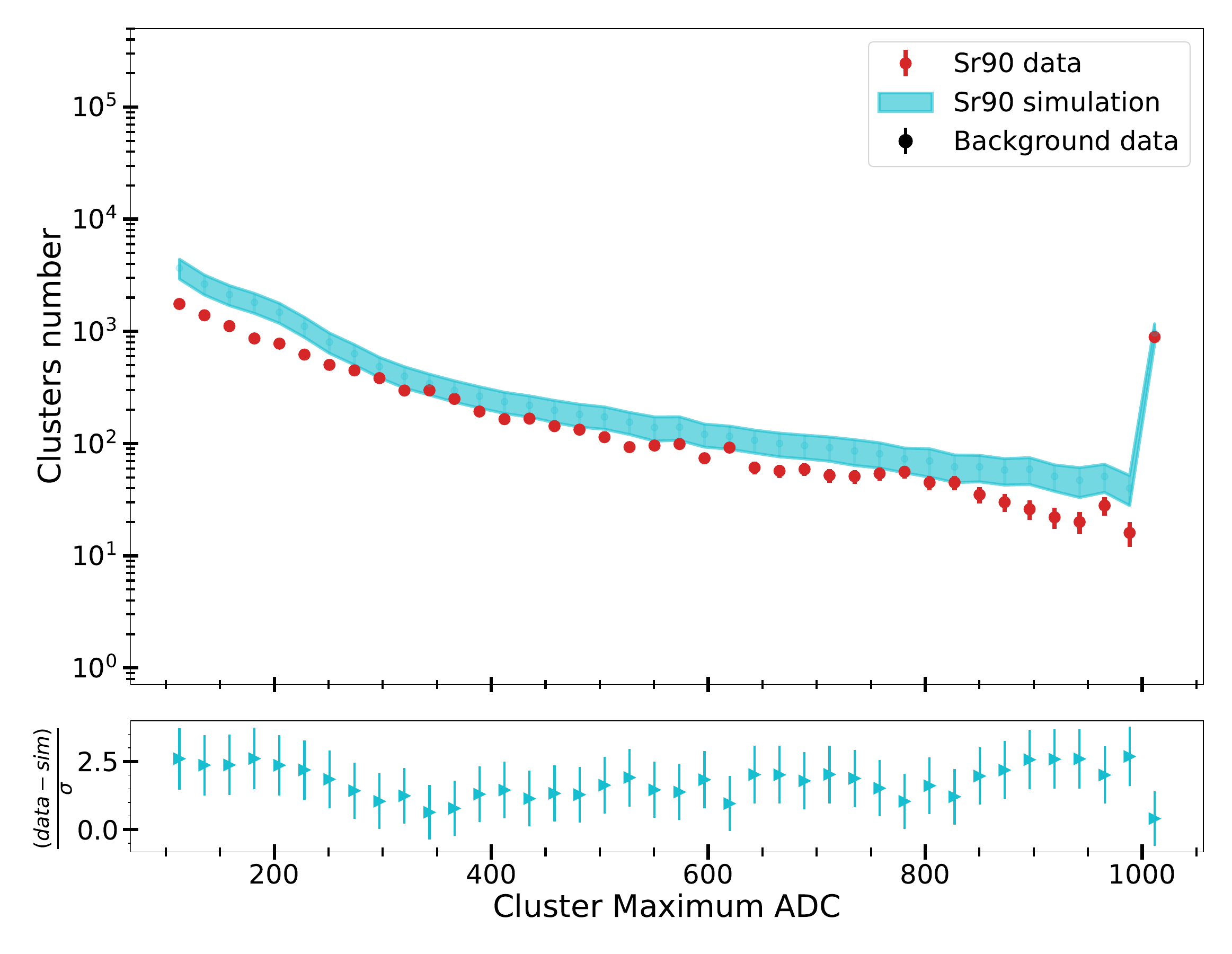}}
{\includegraphics[width=0.47\textwidth]{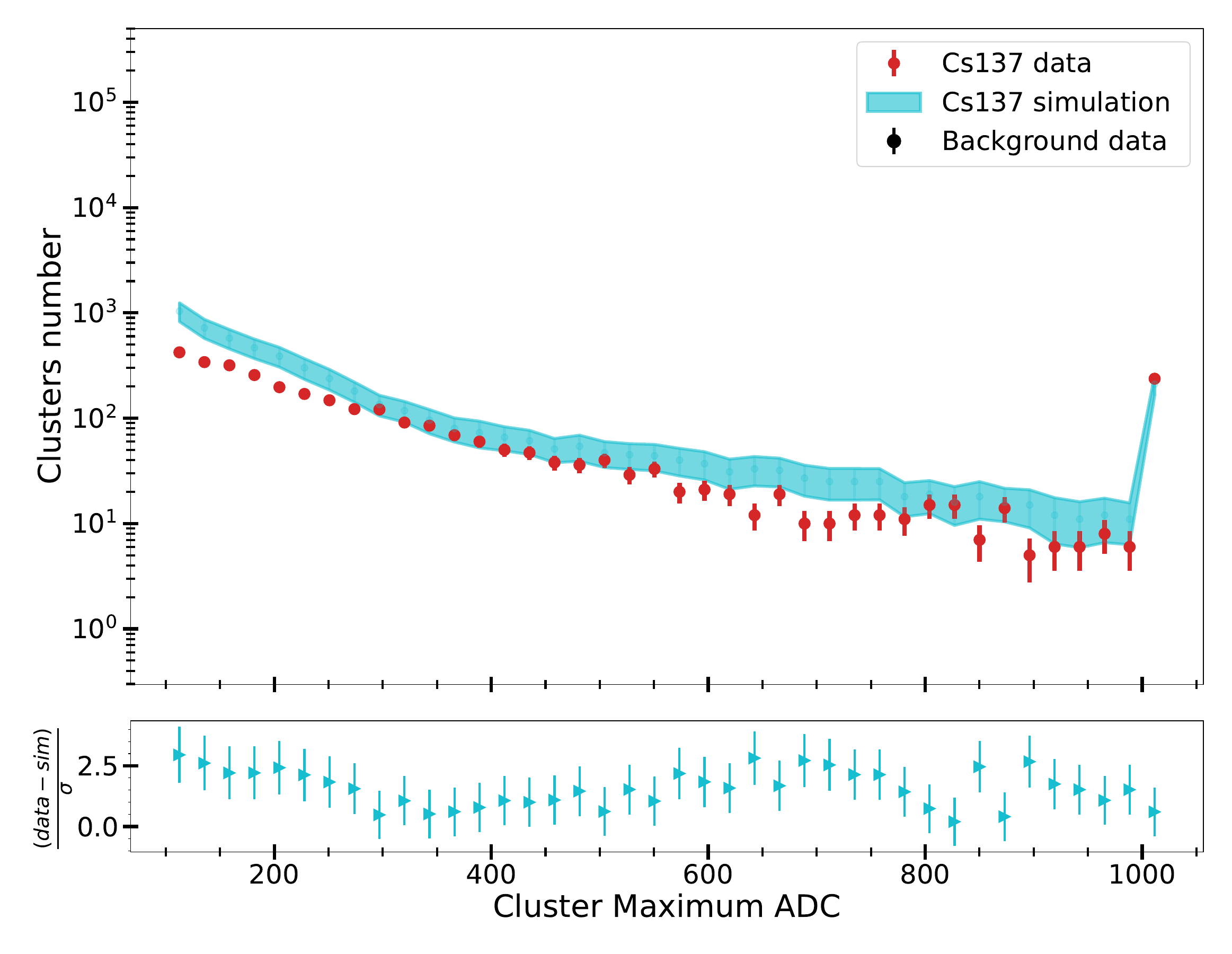}}
\caption{
Maximum ADC signal per cluster distribution for the experimental and simulated data for Sr90 (top left) and Cs 137 (top right), the experimental data applying the SNR procedure and comparing with pure simulation. At the bottom are the same distributions applying the CC procedure for Sr90 (bottom left) and Cs 137 (bottom right) at 0 mm between sensor and source for 500 s exposure time.
}
\label{fig:plot_hist_max_ADC_cut_5clst}
\end{figure}

Another parameter we analyze is the distribution of the maximum ADC signal per cluster for the same data set as before, shown in figure \ref{fig:plot_hist_max_ADC_cut_5clst}. In the upper panels, we have the maximum ADC signal per cluster distribution after applying the SNR procedure. In this case, 
the simulated data distribution shows greater efficiency since it is higher than the experimental data along the entire range of maximum ADC signal per cluster.

In addition, we appreciate the signal saturation at a maximum of 1023 ADC. At the bottom panels we apply the SNR and CC procedures. As a result, the $\chi^2_\nu$ for the maximum ADC signal per cluster distributions is 2.62 for Sr90 and 1.94 for Cs137.

To give the reader a summarized perspective of our results, we present in Table \ref{table:chi2_test_clst5_max100}, the $\chi^2_\nu$ for the three observables we study: the ADC distribution, the cluster size, and the maximum ADC per cluster distributions. The SNR filter has been applied to the experimental data for all of these distributions. As mentioned before, an extra cut rejecting pixels with less than or equal to 100 ADC units has been applied for the ADC distributions in the data and the simulation. For the cluster size and the maximum ADC per cluster distributions, as mentioned already, we apply the CC procedure for simulation and data.

As we can see, the best agreement we obtained is for the ADC distributions, being very reasonable the ones we got for the cluster size and the maximum ADC per cluster distributions, considering the limitations we face. In that sense, it is crucial to remark that the design characteristics of the OV5647 CMOS are aimed at optimizing a photographic image by artificially amplifying the intensities and other parameters of the pixels. The latter is out of our control, since it cannot be handled via standard software.

\begin{table}[htbp]
\centering
\caption{
The $\chi^2_\nu$ tests with 40 bins when comparing the experimental data applying the SNR procedure with simulation. For the ADC distributions with a cut of 100 ADC, and for the distributions related to the number of clusters (i.e., size, and maximum ADC signal) with CC procedure for Sr90 and Cs137 at 0 mm distance between sensor and source for 500 s exposure time.\label{table:chi2_test_clst5_max100}
}
\smallskip
\begin{tabular}{l|rr}
\hline
& \multicolumn{2}{c}{\textbf{$\chi^2_\nu$}} \\ \cline{2-3}
\textbf{Source} & \textbf{Sr90} & \textbf{Cs137} \\ 
\hline
ADC number (cut >100 ADC) & 0.45 & 0.48 \\ 
Cluster Size & 1.77 & 1.26 \\ 
Maximum cluster ADC & 2.62 & 1.94 \\ 
\hline
\end{tabular}
\end{table}

\begin{figure}[htbp]
\centering
{\includegraphics[width=0.475\textwidth]
{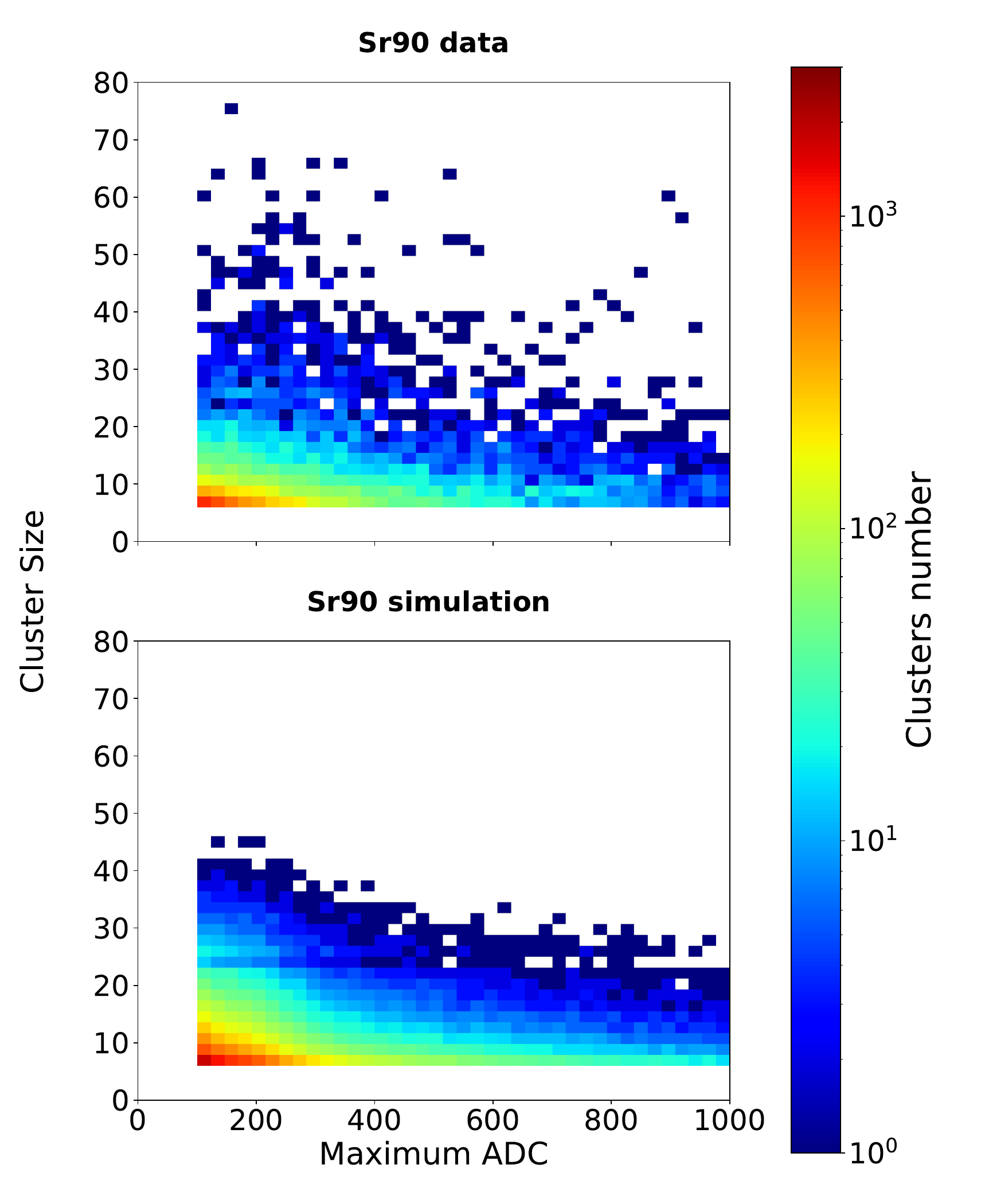}}
{\includegraphics[width=0.475\textwidth]
{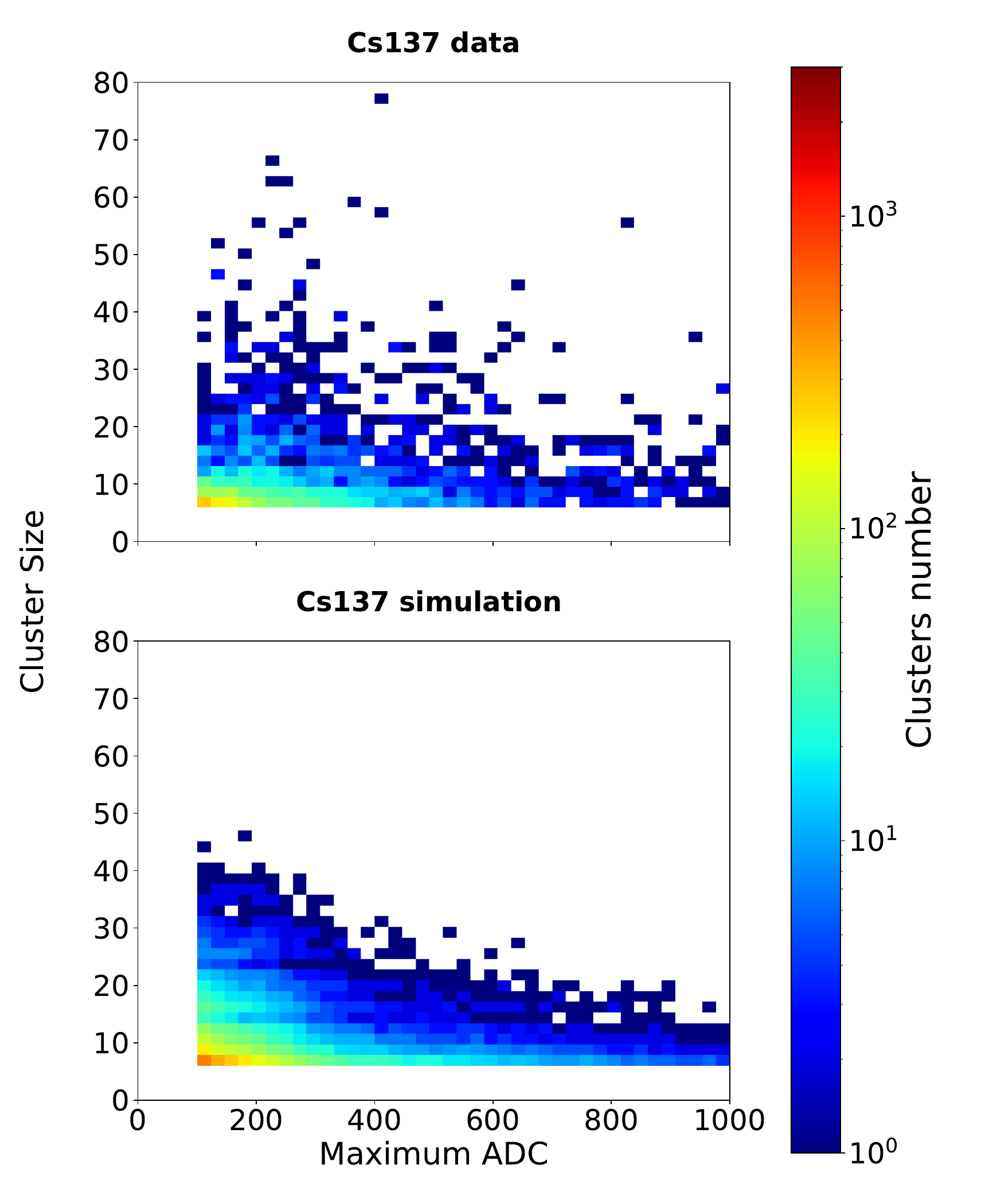}}
\caption{
Cluster size versus maximum ADC signal per cluster for the experimental and simulated data for Sr90 (left) and Cs 137 (right), the experimental data applying the SNR procedure and the CC procedure was applied to experimental data and simulation to compare both, at 0 mm between sensor and source for 500 s exposure time.
}
\label{fig:plot_hists2D_Cluste_vs_Max_clst5_max100}
\end{figure}

We also looked for correlations between variables in 2D histograms of the cluster size vs maximum ADC signal per cluster, for Sr90 and Cs137 at 0 mm distance between source and detector for the data and the simulation. Fig \ref{fig:plot_hists2D_Cluste_vs_Max_clst5_max100} show these distributions.

In figure \ref{fig:plot_hists2D_Cluste_vs_Max_clst5_max100}, we can observe that the simulated data for these distributions are close to the experimental data for both Sr90 and Cs137. Due to the background noise, the experimental data show more scattered clusters above 30 pixels than in the simulation. The $\chi^2_\nu$ for these distributions is 0.66 for Sr90 and 0.44 for Cs137, showing a good agreement. 

\begin{figure}[htbp]
\centering
\includegraphics[width=0.5\textwidth]{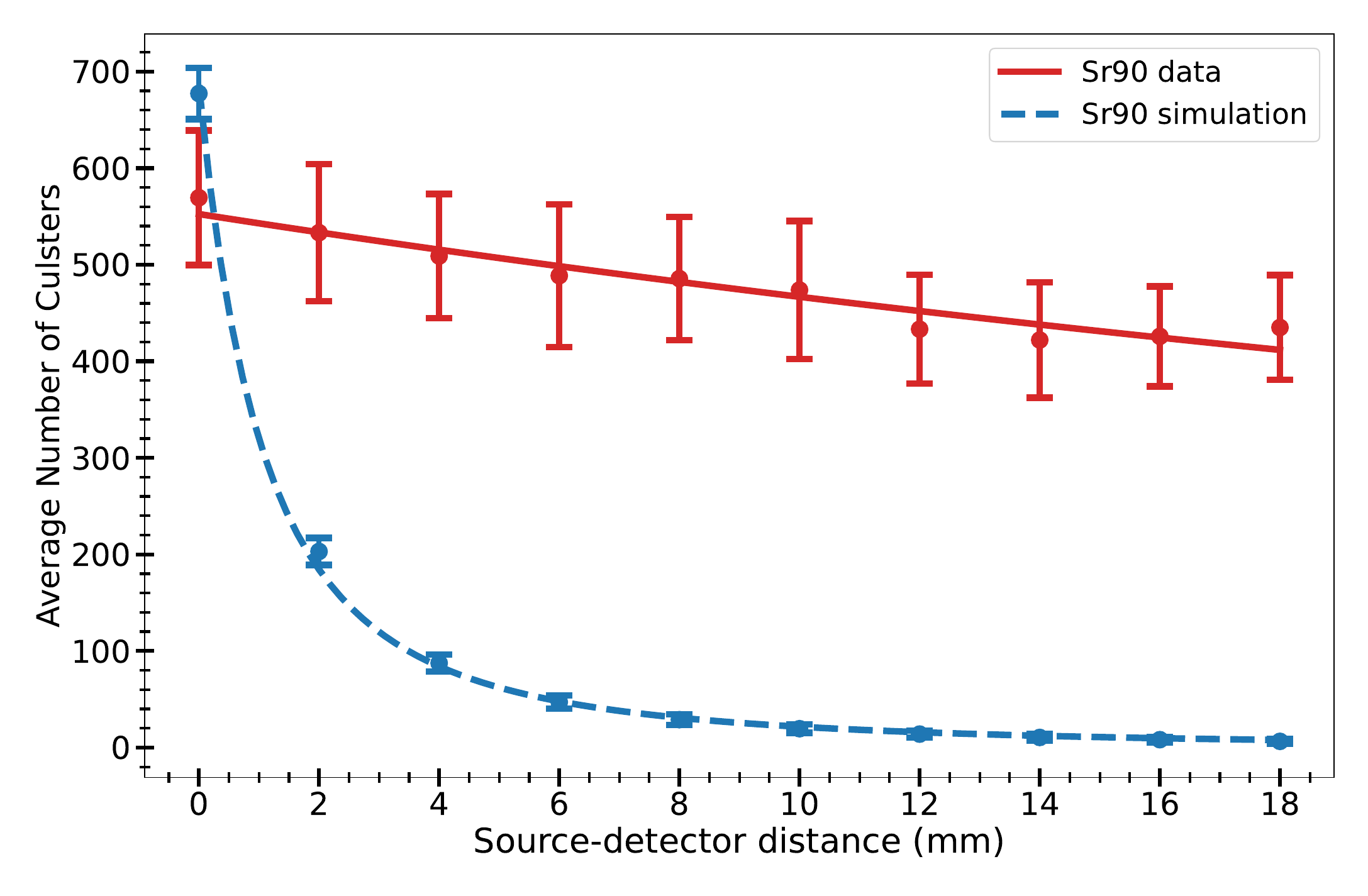}
\caption{
Average cluster number, as a function of distance ($z$) for the Sr90 source. The experimental data applying the SNR procedure is in red and the pure simulation without cuts is in blue. Both data and simulation are fitted to the inverse square function $y=a/(z+b)^2$.
}
\label{fig:plot_fit_inv2_sr90_nocut}
\end{figure}

In addition, to verify the inverse square distance law behavior of the Sr90 radioactive source we performed measurements every 2 mm from 0 to 18 mm between the source-detector distance, taking 100 frames per each distance selected. In figure \ref{fig:plot_fit_inv2_sr90_nocut} we show the average number of clusters at different distances ($z$) for both the experimental data with noise reduction SNR procedure and the simulation and their respective fits to $a/(z+b)^2$ (i.e. the inverse square of the distance, where $a$ and $b$ are free parameters)

The simulation shows the expected behavior with distance, but the experimental data does not. This is because there is a large background noise of a few pixel clusters, as mentioned previously, due to the physical and acquisition software characteristics of the CMOS sensor. As the source moves away from the sensor, the background noise remains constant while the signal clusters diminish, showing a more linear behavior for the experimental data.

\begin{figure}[htbp]
\centering
\includegraphics[width=0.5\textwidth]{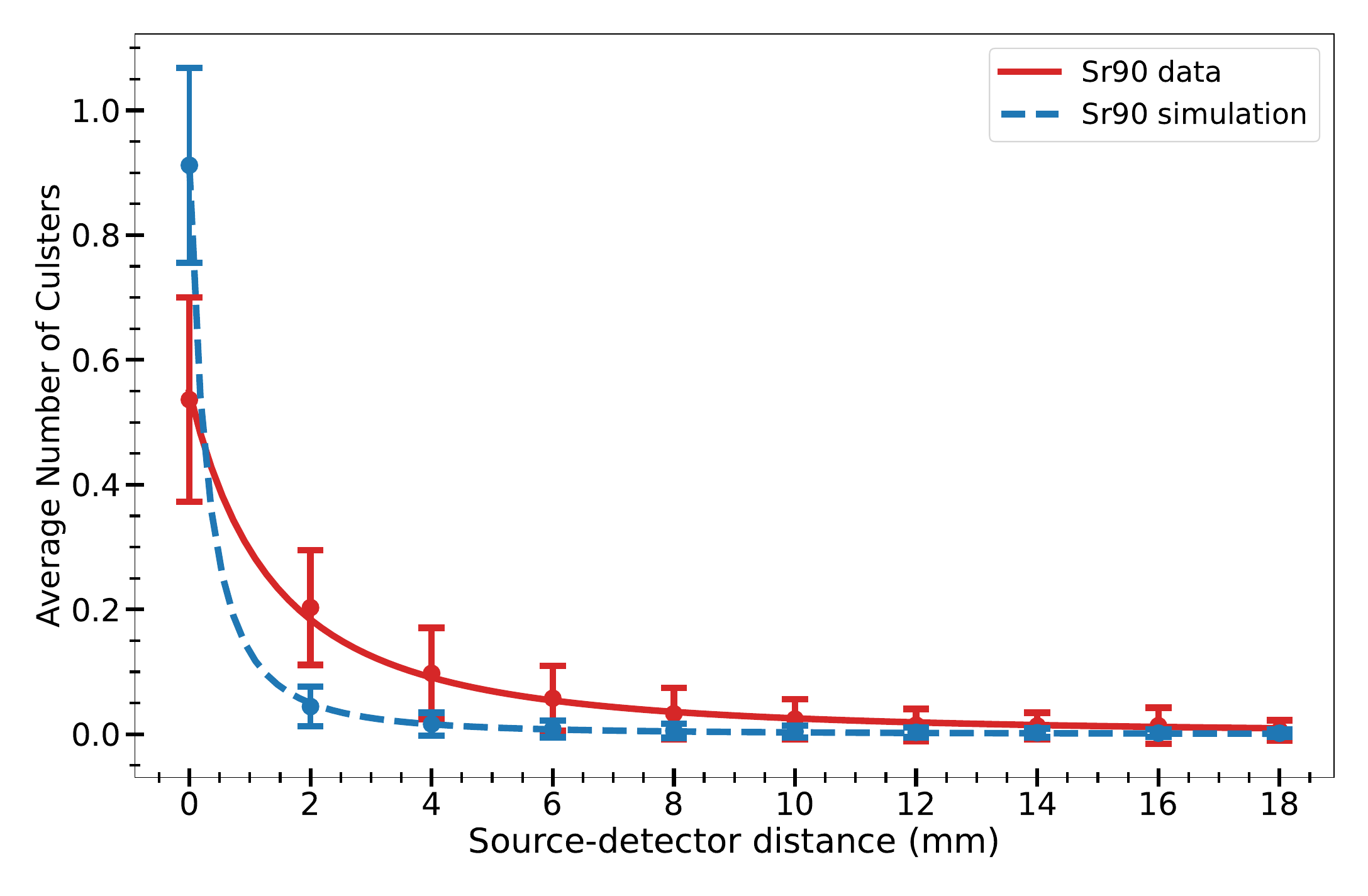}
\caption{
Average cluster number, as a function of distance ($z$) for the Sr90 source. The SNR procedure was applied to the experimental data, while for both experimental and simulation data the CC procedure was applied. The experimental data is in red and the simulation data is in blue. Both data are fitted to the inverse square function $y=a/(z+b)^2$.
}
\label{fig:plot_fit_inv2_sr90_datcut_simnocut}
\end{figure}

To show the expected behavior of the signal with distance, we apply the SNR procedure to the experimental data, while for both experimental and simulation data the CC procedure was applied, to remove the background. We normalize both distributions to better appreciate their behavior with the source-detector distance as the cuts affect the average cluster count. 
Figure \ref{fig:plot_fit_inv2_sr90_datcut_simnocut} shows that these adjustments follow the inverse square of the distance. The parameters obtained for these fits are $a = 0.16 \pm 0.02$, $b = 0.41 \pm 0.03$ for the experimental data, and for the simulation $a = 0.34 \pm 0.03$, $b = 0.61 \pm 0.02$, with a $\chi^2_\nu$ of 0.018 and 0.014 for the fit of experimental and simulated data, respectively. This demonstrates the expected behavior for the source-detector distance for both experimental data and the simulation. Comparing both fits we get a $\chi^2_\nu$ of 0.98 this shows that both fits represent the same behavior with the distance.

\begin{figure}[htbp]
\centering
{\includegraphics[width=0.47\textwidth]{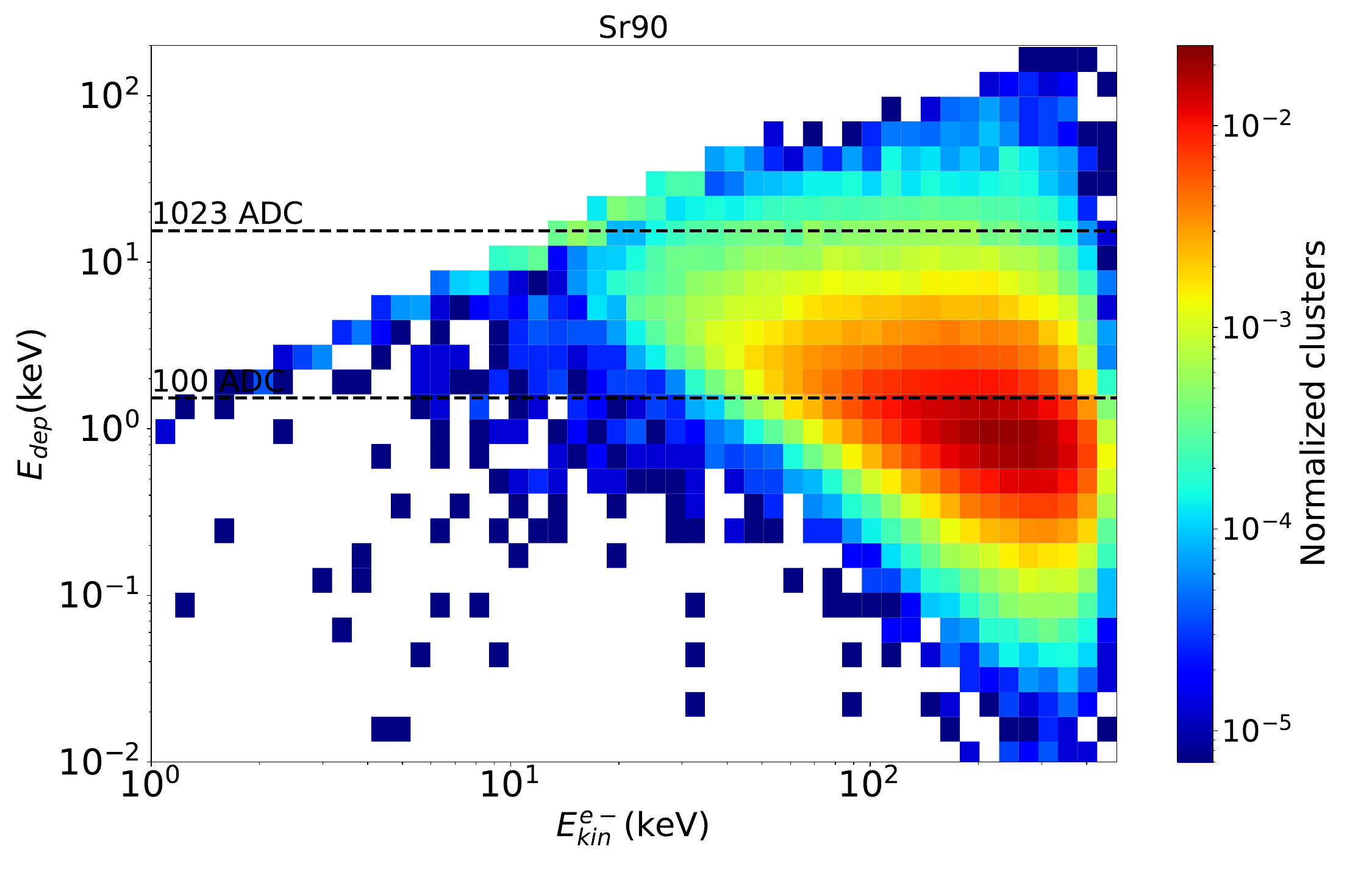}}
{\includegraphics[width=0.47\textwidth]{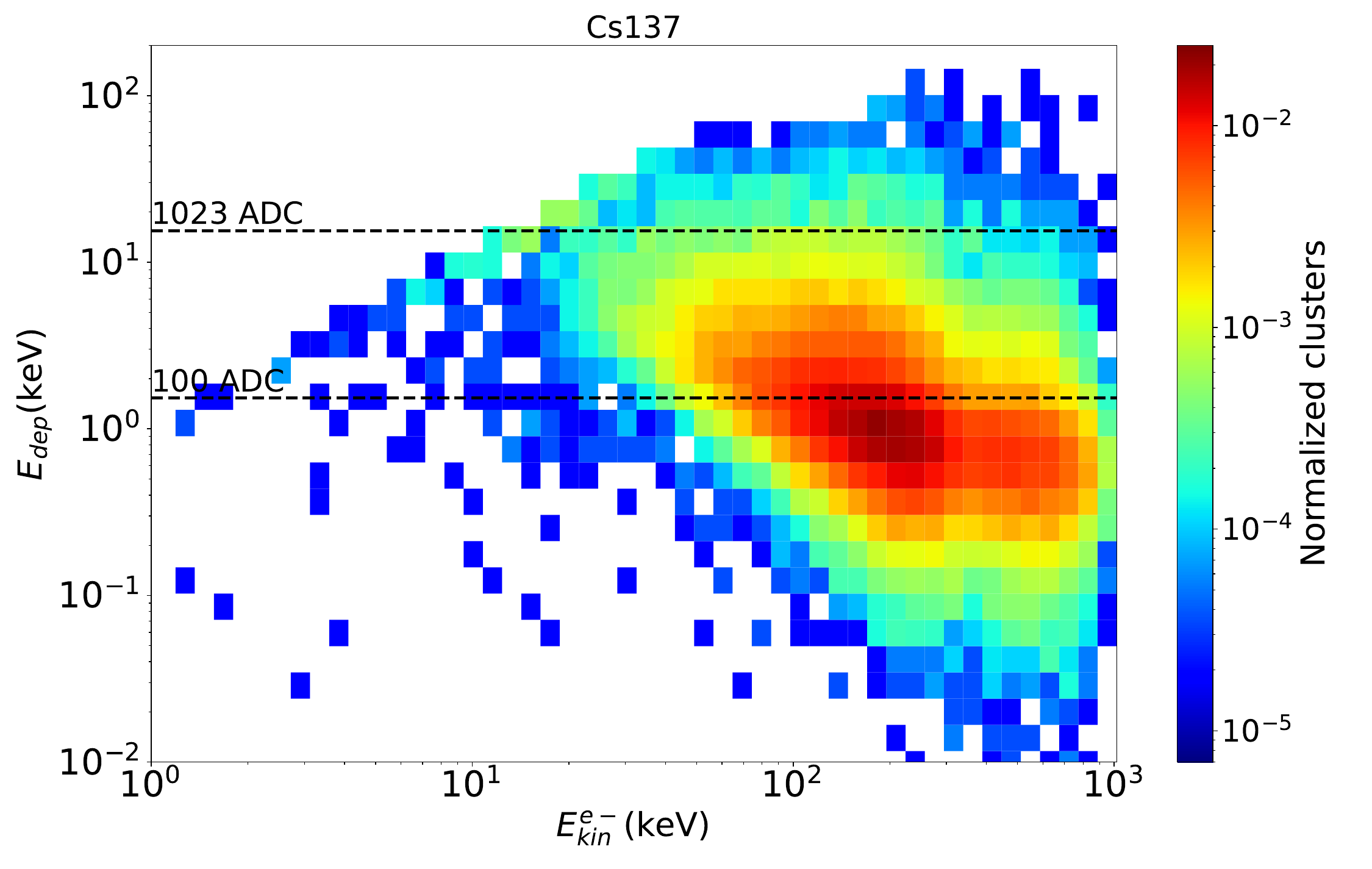}}
{\includegraphics[width=0.47\textwidth]{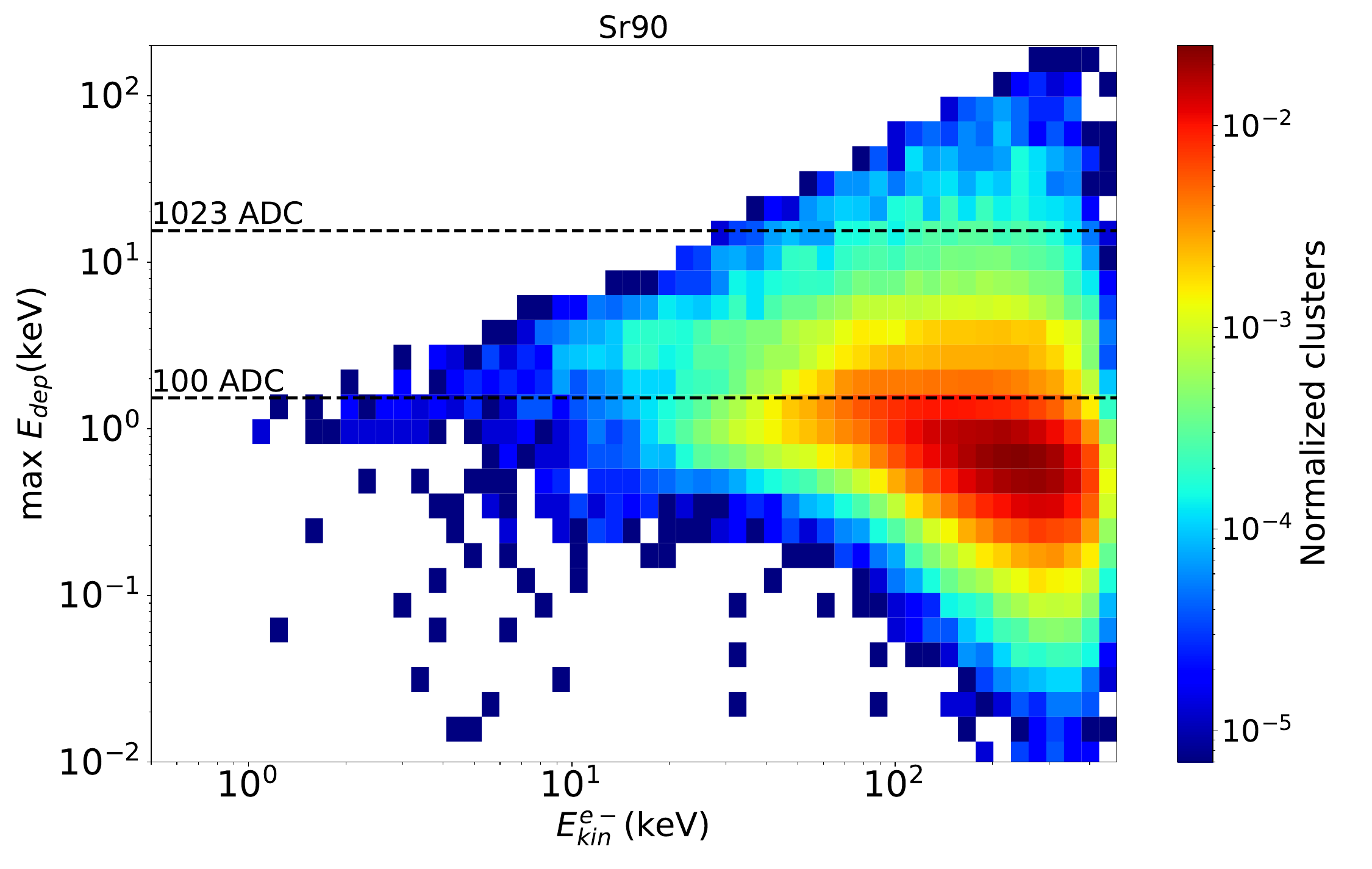}}
{\includegraphics[width=0.47\textwidth]{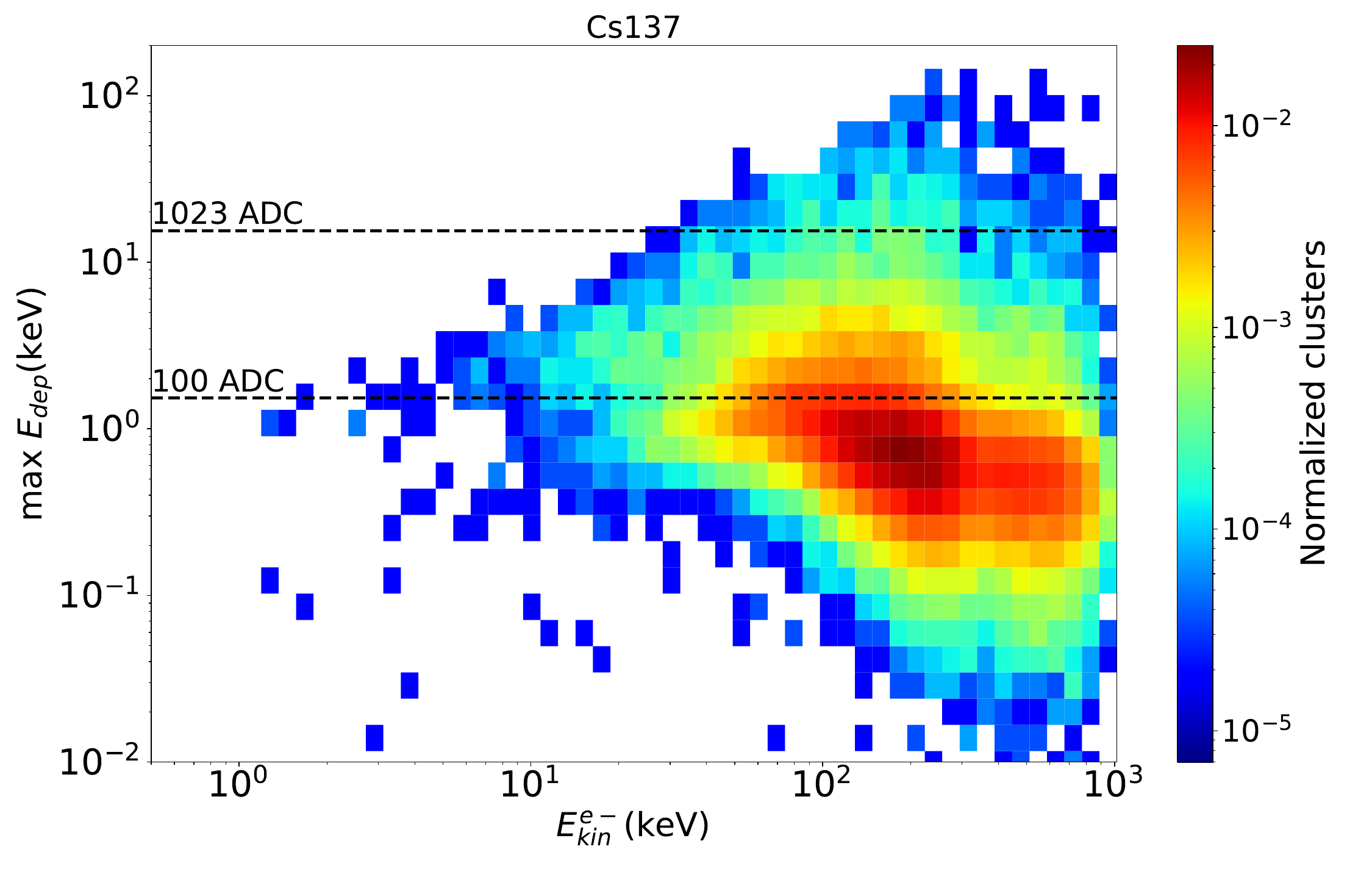}}
{\includegraphics[width=0.47\textwidth]{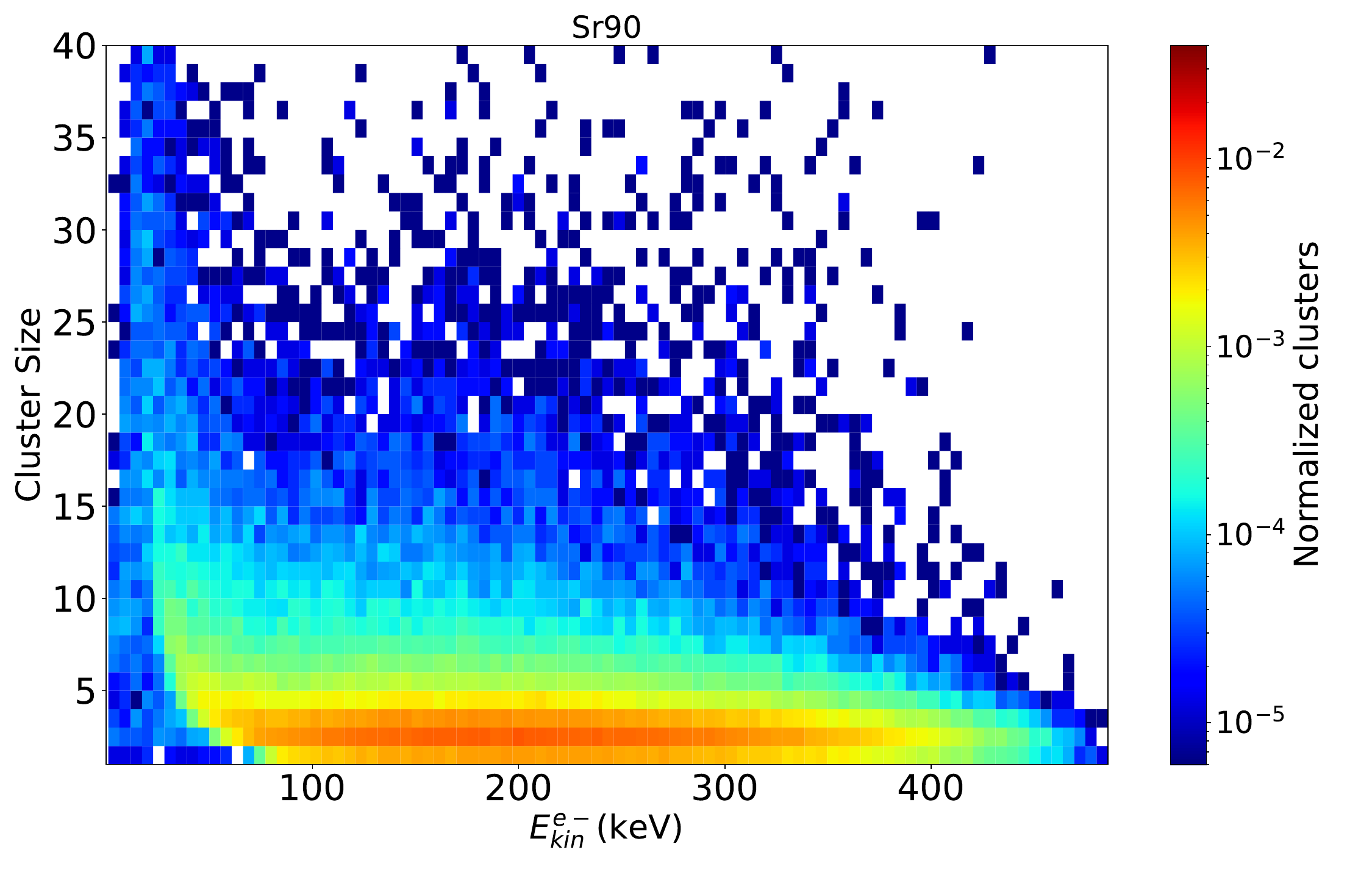}}
{\includegraphics[width=0.47\textwidth]{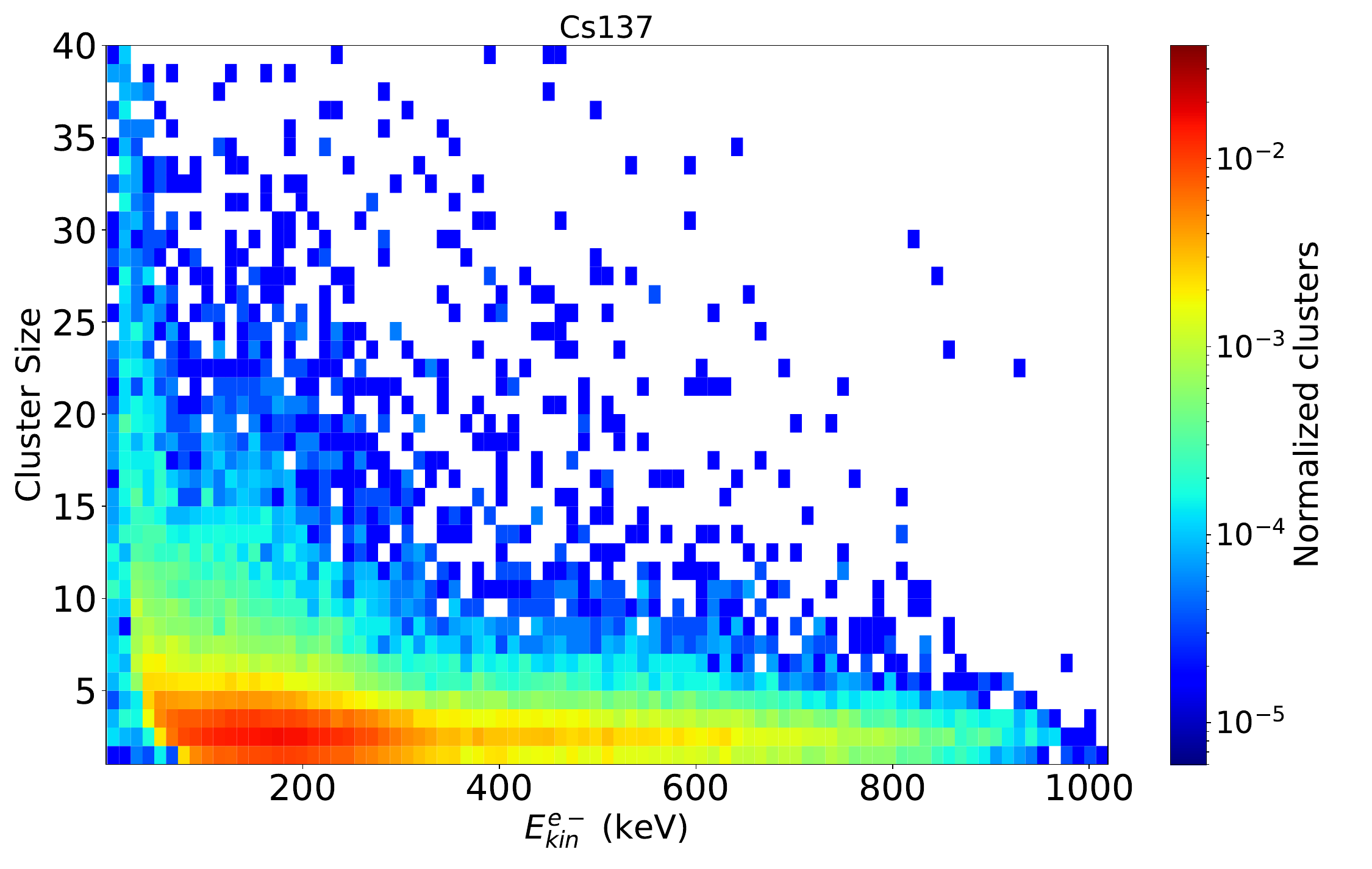}}
\caption{Geant4 simulation of the true kinetic energy of arriving electrons at the silicon pixels emitted by Sr90 (left) and Cs137 (right) versus: a) the total energy deposited in the cluster of pixels, b) the maximum energy deposited in the cluster c) the size of the cluster. The dashed lines are located in the energy corresponding to the ADC lower cut and saturation. 
}
\label{fig:energyresolution}
\end{figure}

To explore the energy resolution capabilities of the commercial CMOS sensor, we used Geant4 simulations to analyze the correlation between the primary electron energy at the point of interaction with the silicon pixel and three parameters: (a) the total energy deposited in the cluster, (b) the maximum energy deposited in a single pixel within a cluster, and (c) the cluster size. The resulting correlations for both radioactive sources are presented in figure \ref{fig:energyresolution}. No parameter exhibits a clear correlation with the primary electron energy. Within the energy range sensitive to the sensor, delimited by the horizontal black dashed lines, we observe an almost flat response in both energy deposition related parameters with respect to the incident electron's kinetic energy. This behaviour is even stronger for the cluster size, where small clusters can be produced by both low-energy and high-energy electrons. Interestingly, for very low deposited energies, a slight anticorrelation trend with the primary electron's kinetic energy is observed. However, this region is shadowed by background contributions and is therefore not accessible for a reliable analysis. 

\begin{figure}[htbp]
\centering
{\includegraphics[width=0.6\textwidth]{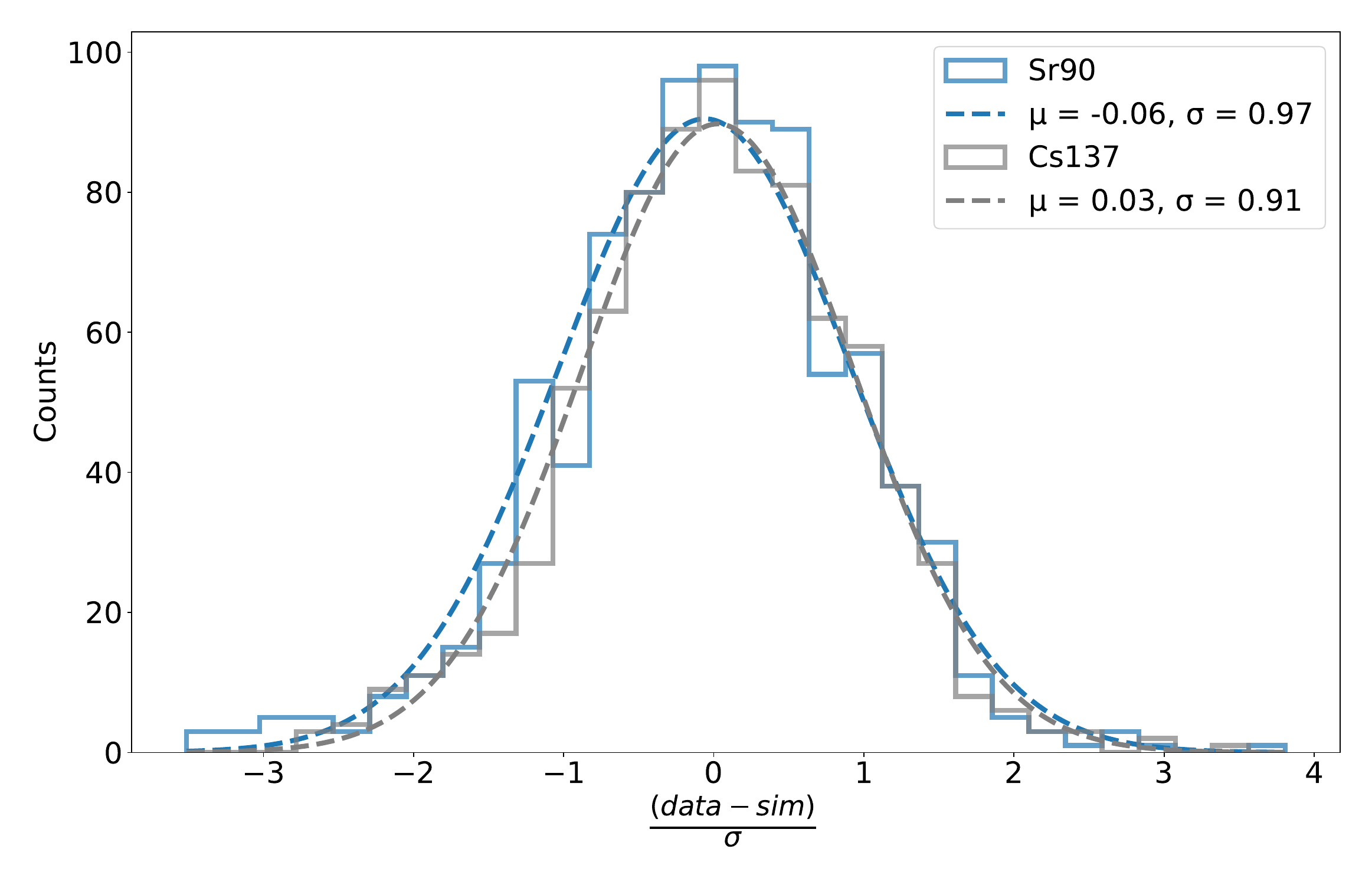}}
\caption{Distributions of the differences in counts per ADC bin (i.e. deposited energy above 100 ADC) between experimental data and simulation, expressed in units of standard deviations (sigmas), for both Sr90 and Cs137. Each distribution is fitted with a Gaussian function. The fit parameters are shown in the legend. 
}
\label{fig:depEner_comp}
\end{figure}

Even if the primary energy cannot be estimated, the sensor setup is capable of accurately describing the energy deposited in the pixel matrix in terms of ADC counts. As seen in figure \ref{fig:plot_hist_ADC_cut} there was a good agreement between simulation and experimental data. To study this agreement in detail, we compared in figure \ref{fig:depEner_comp} for Sr90 and Cs137 the counts in bins of 1 ADC, greater than 100 ADC, between experimental data and simulation, normalized by the standard deviation. After fitting both distributions with a Gaussian function, we observe that the errors are almost symmetrical centered around zero, with a sigma of less than one standard deviation. 

\begin{figure}[htbp]
\centering
{\includegraphics[width=0.47\textwidth]{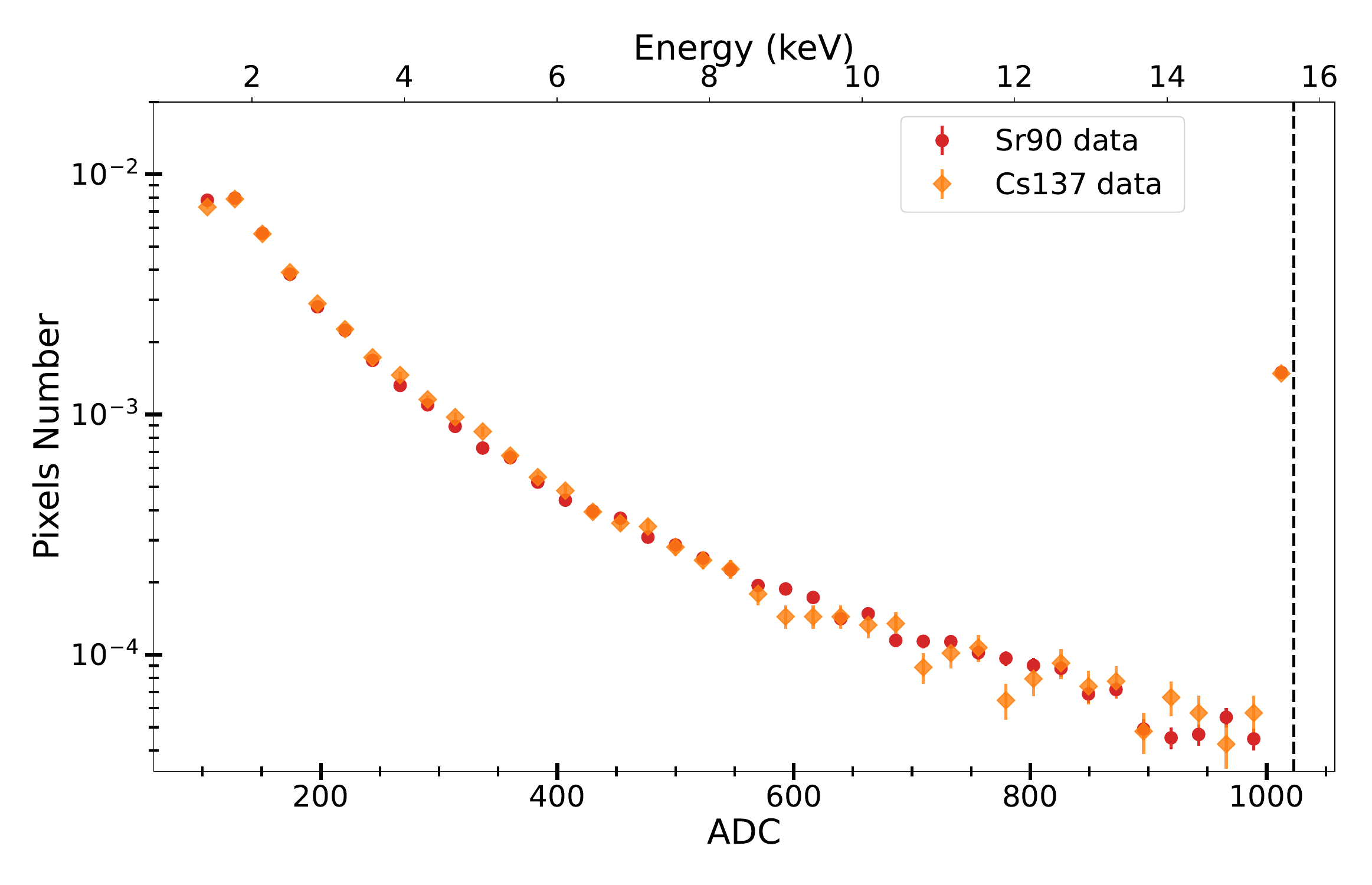}}
{\includegraphics[width=0.47\textwidth]{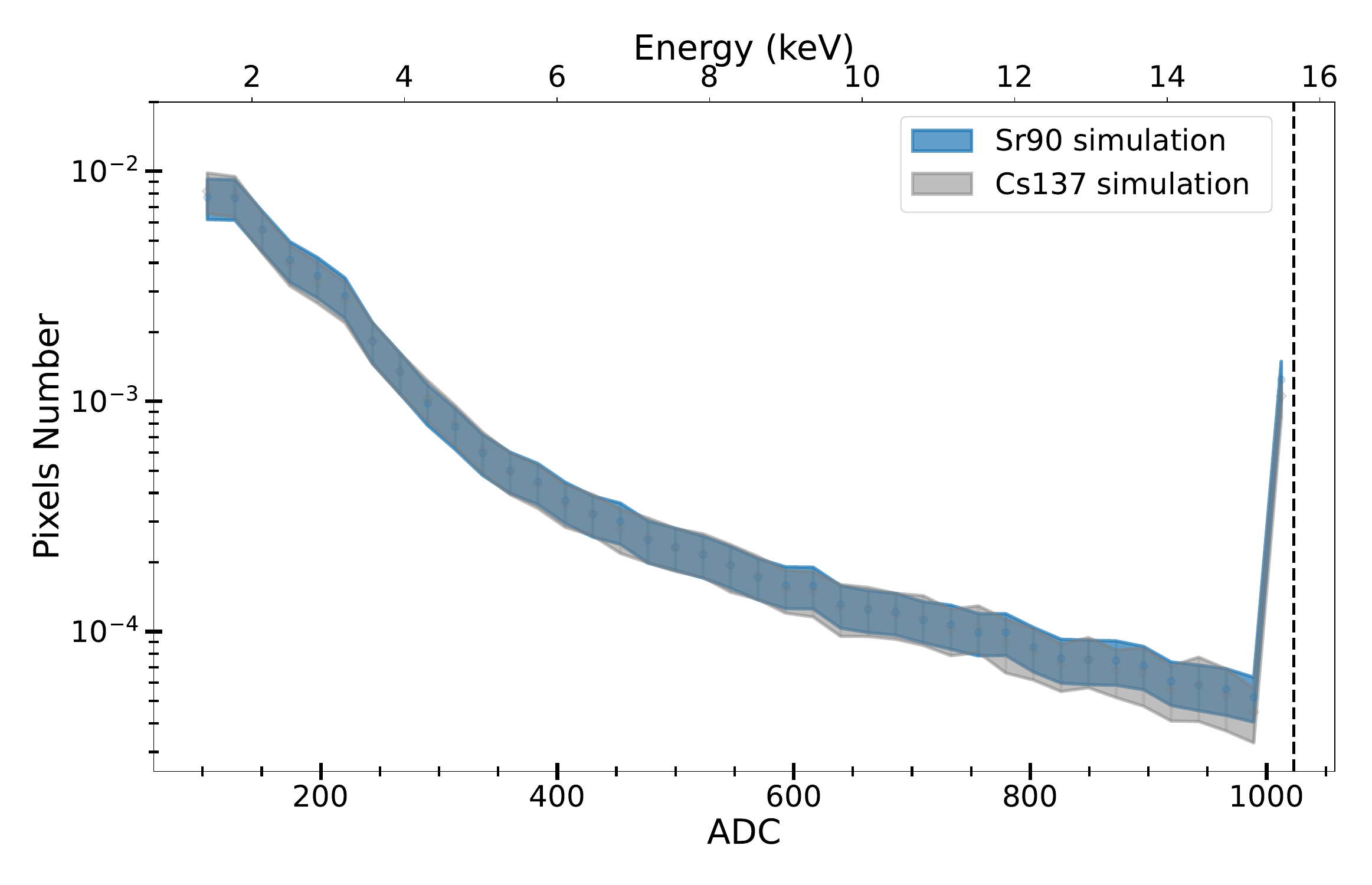}}
\caption{
Normalized ADC number distribution comparing the sources of Sr90 and Cs137 for the experimental data (left) and the simulation (right), experimental data applying the SNR procedure comparing with simulation applying a cut to both to remove values less than or equal to 100 ADC to both the experimental data and the simulation at 0mm between sensor and source, for 500 s exposure time. 
}
\label{fig:plot_hist_ADC_comp_sr_cs_obs_sim_norm}
\end{figure}

\begin{figure}[htbp]
\centering
{\includegraphics[width=0.47\textwidth]{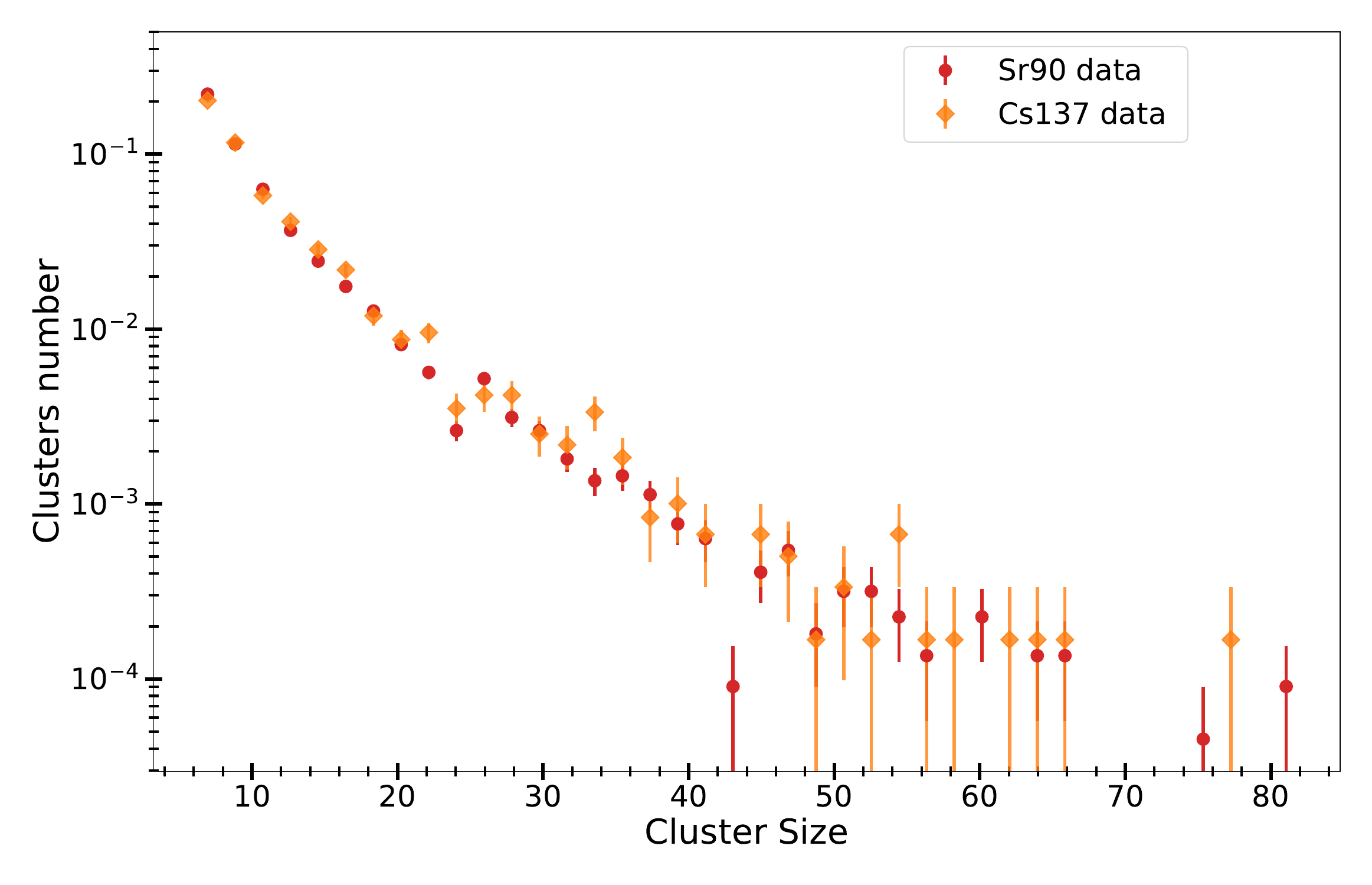}}
{\includegraphics[width=0.47\textwidth]{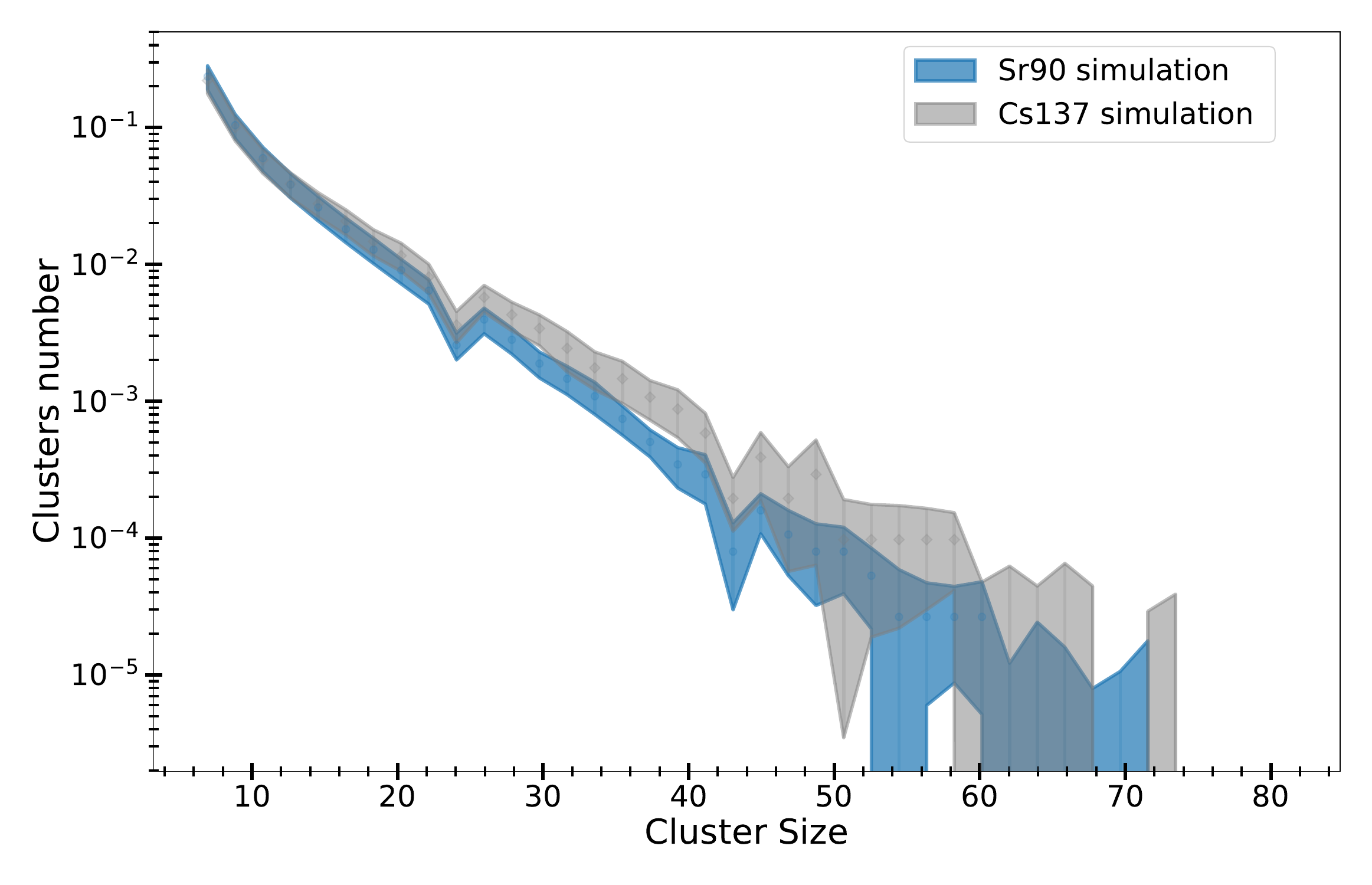}}
\caption{
Normalized cluster size distribution comparing the sources of Sr90 and Cs137 for the experimental data (left) and the simulation (right), experimental data applying the SNR procedure comparing with simulation, and applying the CC procedure to both at 0mm between sensor and source for 500 s exposure time.
}
\label{fig:plot_hist_cluster_comp_sr_cs_obs_sim_norm}
\end{figure}

\begin{figure}[htbp]
\centering
{\includegraphics[width=0.47\textwidth]{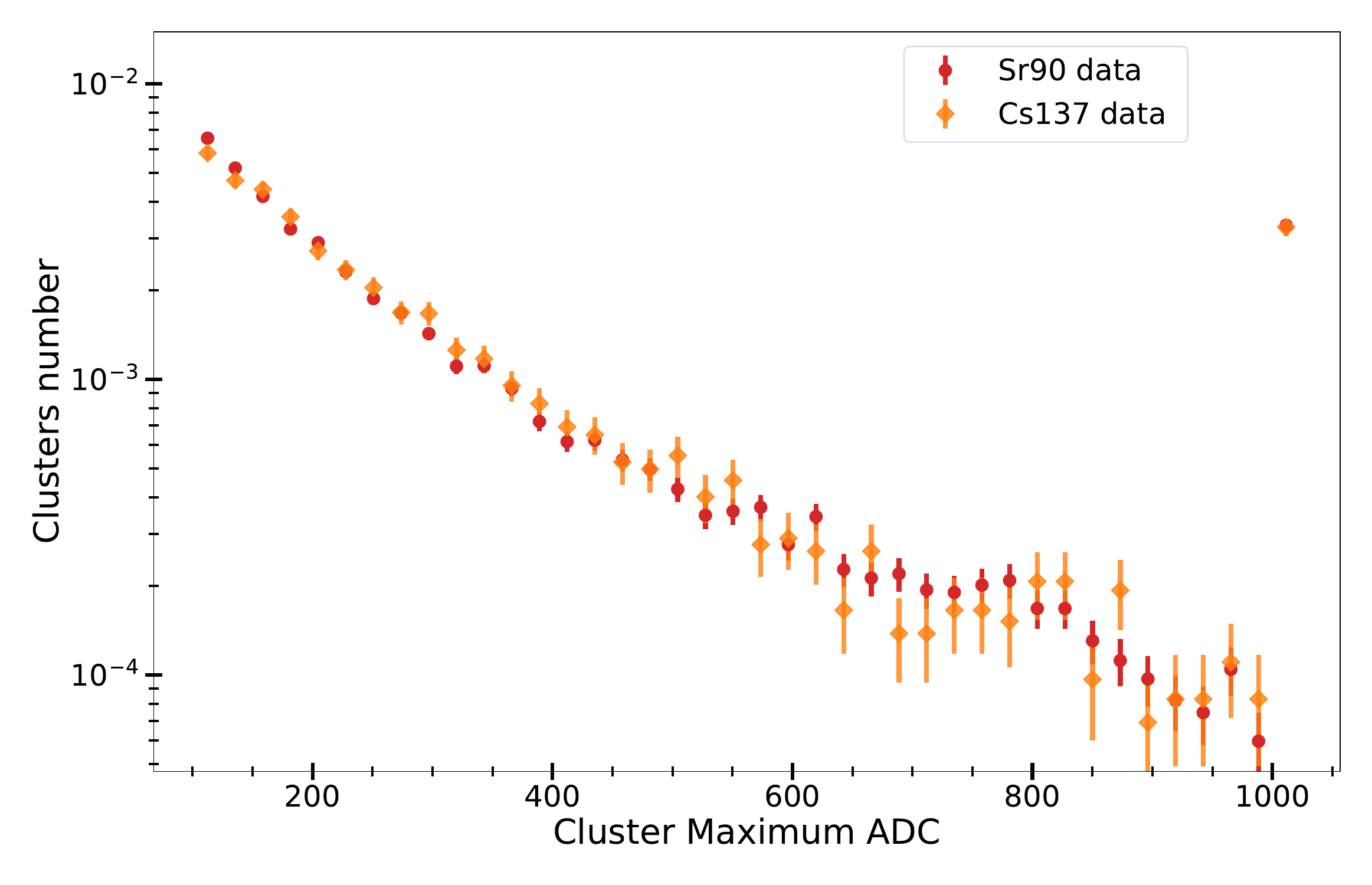}}
{\includegraphics[width=0.47\textwidth]{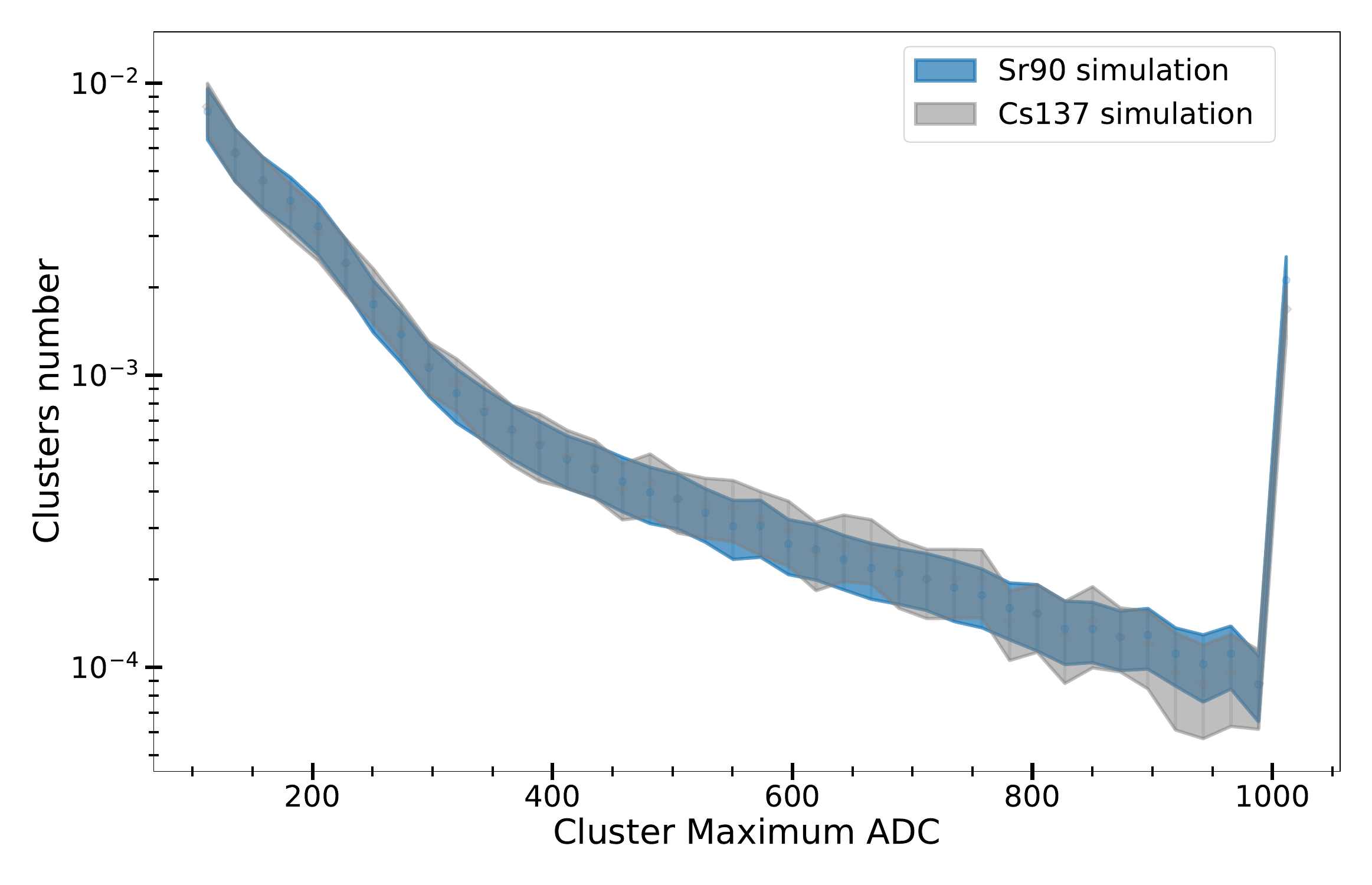}}
\caption{
Normalized maximum ADC signal per cluster distribution comparing the sources of Sr90 and Cs137 for the experimental data (left) and the simulation (right), experimental data applying the SNR procedure comparing with simulation, and applying the CC procedure to both at 0mm between sensor and source for 500 s exposure time.
}
\label{fig:plot_hist_max_ADC_comp_sr_cs_obs_sim_norm}
\end{figure}

Given the poor energy resolution of the sensor and following what was seen in figure \ref{fig:emitted_deposited}, the sensor has no radioactive source separation power. However, to further test whether the Sr90 and Cs137 sources can be distinguished, we compared the distributions discussed above at the 0-mm distance. Then, we normalize the histograms to avoid differences in the activities of the sources. Figure \ref{fig:plot_hist_ADC_comp_sr_cs_obs_sim_norm}, \ref{fig:plot_hist_cluster_comp_sr_cs_obs_sim_norm}, and \ref{fig:plot_hist_max_ADC_comp_sr_cs_obs_sim_norm} show the normalized distributions at the pixel and cluster level after applying the cuts. The $\chi^2_\nu$ for the comparison between the two sources in the experiment and the simulation is shown in Table \ref{table:chi2_test_comp}. The average $\chi^2_\nu$ for the experimental data is 1.33, and for the simulation, it is 0.25. Thus, it is impossible to distinguish between these two radioactive sources for the camera capabilities, even if the emitted decay radiation spectrum are different. 

\begin{table}[htbp]
\centering
\caption{
The $\chi^2_\nu$ test with 40 bins when comparing the experimental data applying the SNR procedure, and simulation of Sr90 with Cs137 for the normalized ADC distributions with a cut of 100 ADC and the normalized distributions related to the number of clusters (i.e. size, and maximum ADC signal) with the CC procedure at 0 mm distance between source and detector for 500 s exposure time.
\label{table:chi2_test_comp}
}
\smallskip
\begin{tabular}{l|cc}
\hline
& \multicolumn{2}{c}{\textbf{$\chi^2_\nu$}} \\ \cline{2-3}
\textbf{Source} & \textbf{Experiment} & \textbf{Simulation} \\ 
\hline
ADC number (cut >100 ADC) & 1.72 & 0.03 \\ 
Cluster Size & 1.39 & 0.66 \\ 
Maximum cluster ADC & 0.89 & 0.06 \\ 
\hline
\end{tabular}
\end{table}

\section{Conclusions} \label{conclu}

In this work, we implement a system for the data acquisition of the OmniVision OV5647 CMOS sensor with the help of a Raspberry Pi. We give precise information about the settings needed for the camera to take stable and sensitive frames for detecting particles. We define a background reduction procedure, to eliminate experimental noise. We first subtract the average ADC value per pixel and apply a 5 sigma ADC threshold per pixel obtaining a reduction of 99.995\% of active pixels. Then we remove clusters with a size of 5 pixels or fewer and clusters where one or more of its pixels is less than or equal to 100 ADC.

We conclude from the $\chi^2_\nu$ of the analyzed distributions, for Sr90 and Cs137, that the customized Geant4 simulation is in good agreement with the experimental data, considering the characteristics of the CMOS sensor. It is possible to carry out simulations of commercial sensors such as CMOS cameras if their physical specifications are precisely described. In addition, we can reproduce the correlation between variables (e.g. cluster size vs. maximum ADC per cluster) with a $\chi^2_\nu$ that shows good agreement for Sr90 and Cs137. Also, we find that both the experimental data and the simulation follow an inverse square distribution with the distance when the background noise is completely removed, as expected. Thus, this simulation, which has been cross-checked with data, can be used to test the feasibility of further particle detection ideas without the need to implement an experimental setup. 

When comparing the experimental data for the Sr90 and Cs137 sources using the different analyzed parameters, we find that it is not possible to distinguish between the two radioactive sources with this experimental approach, nor through simulation alone. This limitation is related to the poor energy resolution of the sensor, which arises from the lack of a clear correlation between the deposited energy and the primary electron energy. However, given the accurate measurement of energy deposition corroborated by the simulation estimation, we have demonstrated that the sensor, once calibrated, is suitable for dosimetric measurements of source activities.

A refinement of this work \footnote{M. Bonnett Del Alamo et al. work in preparation.} involves applying machine learning techniques to distinguish background clusters from signal ones. This approach could achieve comparable background rejection without sacrificing signal efficiency. Consequently, the ADC background cut could be loosened, which could enhance the sensor’s energy resolution, since in the low energy deposition region there is an anti-correlation between primary and deposited energy. Furthermore, the same method could be used for particle identification by analyzing cluster shapes and comparing them between data and simulation.


\acknowledgments

The authors gratefully acknowledge the Direccion de Gestion de la Investigacion (DGI-PUCP) for funding under grant DGI-2021-C-0020/PI0758. C.S. acknowledges support from CONCYTEC, scholarship under grant 236-2015-FONDECyT. We would also wish to thank José Lipovetzky and Xavier Bertou for the information on the physical characteristics of the Omnivision OV5647 sensor.
\\

\noindent\textbf{Code Availability Statement.} This article has associated code in a code repository. Available at:  
\url{https://github.com/mbonnett/geant4_python_script_CMOS_OV5647}.



\bibliographystyle{JHEP} 
\bibliography{biblio.bib}






\end{document}